\documentclass[aps,twocolumn,preprintnumbers,showpacs,showkeys]{revtex4}
\usepackage{epsfig}
\usepackage{amssymb,amsmath,amsfonts,amsthm,graphicx}
\graphicspath{{./Figures/}}                 
\newcommand{\noi}{\noindent}
\newcommand{\beq}{\begin{equation}}
\newcommand{\eeq}{\end{equation}}
\newcommand{\bea}{\begin{array}}
\newcommand{\eea}{\end{array}}
\newcommand{\beqa}{\begin{eqnarray}}
\newcommand{\eeqa}{\end{eqnarray}}
\newcommand{\Fig}[1]{Fig.~\ref{#1}}

\newcommand{\Sec}[1]{Section~\ref{#1}}
\newcommand{\Eq}[1]{Eq.~(\ref{#1})}

\def\beqa{\begin{eqnarray}}
\def\eeqa{\end{eqnarray}}
\def\pl{{{\cal P}_\infty}}

\def\Tr{{\rm Tr}}

\begin{document}

\preprint{ITEP-LAT-2014-15,~HU-EP-14/36 }

\title{Dyon structures in the deconfinement phase of lattice gluodynamics: \\
topological clusters, holonomies and Abelian monopoles}

\author{V.~G.~Bornyakov}
\affiliation{Institute for High Energy Physics, 142 281 Protvino, Russia \\
Institute of Theoretical and Experimental Physics, 117259 Moscow, Russia\\
School of Biomedicine, Far Eastern Federal University, 690950,
Vladivostok, Russia }

\author{E.-M.~Ilgenfritz}
\affiliation{Joint Institute for Nuclear Research, VBLHEP and BLTP, 141980 Dubna, Russia}

\author{B.~V.~Martemyanov}
\affiliation{
Institute of Theoretical and Experimental Physics, 117259 Moscow, Russia\\
National Research Nuclear University MEPhI, 115409, Moscow, Russia\\
Moscow Institute of Physics and Technology, 141700, Dolgoprudny, Moscow Region,
Russia}

\author{M.~M\"uller-Preussker}
\affiliation{Humboldt-Universit\"at zu Berlin, Institut f\"ur Physik,
12489 Berlin, Germany}

\date{October 14, 2014}

\begin{abstract}
The topological structure of lattice gluodynamics is studied at
intermediate resolution scale in the deconfining phase with the
help of a cluster analysis.
UV filtered topological charge densities are determined
from a fixed number of low-lying eigenmodes
of the overlap Dirac operator with three types of temporal boundary
conditions applied to the valence quark fields. This method usually
allows to find all three distinguished (anti)dyon constituents
in the gauge field of Kraan-van Baal-Lee-Lu (anti)caloron solutions.
The clustering of the three topological charge densities in Monte Carlo
generated configurations is then used to mark the positions of anticipated
(anti)dyons of the corresponding type. In order to support this
interpretation, inside these clusters, we search also for time-like
Abelian monopole currents (defined in the maximally Abelian gauge) as well
as for local holonomies with at least two approximately degenerated eigenvalues.
Our results support the view that light dyon-antidyon pairs - in contrast to
the heavy (anti)caloron dyon constituents - contribute dominantly to
thermal Yang-Mills fields in the deconfinement phase.

This paper is dedicated to the memory of
Pierre van Baal and Dmitri Igorevich Diakonov who have influenced our work
very much.
\end{abstract}



\maketitle

\section{Introduction}
\label{sec:introduction}

1998 was a remarkable year for lattice gauge theory, in particular
for those interested in chirality and topology. The
Ginsparg-Wilson~\cite{Ginsparg:1981bj} condition to be imposed on  a Dirac
operator and providing a solution of the chirality problem at finite
lattice spacing was rediscovered~\cite{Niedermayer:1998bi},
a concrete construction of the Neuberger overlap Dirac operator was
proposed~\cite{Neuberger:1997fp,Neuberger:1998wv}, and the relation to
topological structure was clarified~\cite{Hasenfratz:1998ri,Niedermayer:1998bi}.

The paradigm of instantons as semiclassical realization of topological
structure at the infrared scale~\cite{Schafer:1996wv} has got a competitor,
when P. van Baal and T.C. Kraan~\cite{Kraan:1998kp,Kraan:1998sn,Kraan:1998pm},
and K.-M. Lee and C.-H. Lu~\cite{Lee:1998bb}, for the case of finite temperature,
worked out a broader class of classical Euclidean solutions of Yang-Mills
theory: calorons with arbitrary holonomy, in the following called
{\it KvBLL calorons}. These solutions have not reached the same level
of acceptance and interest among lattice practitioners
that instantons once had (see, for example ~\cite{Chu:1994vi,Negele:1997bi}).
However, immediate response to the new solutions from the lattice community
can be found in Refs.~\cite{Brower:1998ep,Negele:1998ev}.

Three of us were among the authors of ~\cite{Ilgenfritz:2002qs}
who have demonstrated first that cooling of confining lattice ensembles leads
to the extraction of KvBLL multi-(anti)caloron solutions
(see also ~\cite{Bruckmann:2004nu}).

Shortly later, D. Diakonov, who had been very active before trying to connect
instantons with confinement, in particular by relating the instanton gas to
monopole percolation~\cite{Fukushima:1996mm,Fukushima:1996yn}, wrote his famous
review ``Instantons at work''~\cite{Diakonov:2002fq}, to which he, in a
later version, added a chapter ``Non-instanton semiclassical
configurations''. This extension has become the starting point of a new
research direction. D. Diakonov, together with V. Petrov and other coworkers,
calculated the analog of  't Hooft's instanton amplitude~\cite{Diakonov:2004jn}
and formalized the moduli space of calorons~\cite{Diakonov:2005qa} in terms
of dyon degrees of freedom.

A simulation of a random non-trivial holonomy caloron gas or liquid model
provided already a much better behavior of the potential of a static
quark-antiquark pair towards confinement~\cite{Gerhold:2006sk}
than the corresponding Harrington-Shepard caloron gas model (with trivial
holonomy)~\cite{Harrington:1978ve}.
Surely influenced by Polyakov's work \cite{Polyakov:1976fu} on quark confinement
by monopoles D. Diakonov and V. Petrov have then formulated a dyon
(i.e. monopole) gas model of confinement~\cite{Diakonov:2007nv}, for which
they have been able to present a closed analytical solution (For random monopole
gas simulations see~\cite{Bruckmann:2009nw,Bruckmann:2011yd}.).

In fact, the idea to reformulate the statistical mechanics of a gas of
multi-instantons in terms of the moduli space of their constituents
(``instanton quarks'') had been discussed already much earlier for the
two-dimensional nonlinear sigma model~\cite{Fateev:1979dc,Berg:1979uq}
by transforming the partition function into a Coulomb gas model of the
constituents.

We are aware of recent papers by E. Shuryak and collaborators to
formulate models dealing with the statistical mechanics of selfdual
dyons (and antiselfdual antidyons), partly including the effect of
dynamical fermions~\cite{Shuryak:2012aa,Faccioli:2013ja}. We hope
to come back to this problem. In our present context we want to
refer to Ref.~\cite{Shuryak:2013tka} dealing with the $T$ dependence
of the density of light and heavy ($L$ and $M$) dyons in $SU(2)$
gluodynamics.

During our intensive search for evidence of calorons and dyons in 
Monte Carlo generated lattice gauge field ensembles
\cite{Ilgenfritz:2002qs,Gattringer:2003uq,Ilgenfritz:2004ws,
Ilgenfritz:2004zz,Ilgenfritz:2005um,Ilgenfritz:2006ju}
we came into collaboration with
P. van Baal~\cite{Bruckmann:2004zy,Bruckmann:2004ib}.

In our recent work~\cite{Ilgenfritz:2013oda}, after having searched for
signatures for calorons and dyons for $SU(2)$ Yang-Mills theory by means
of overlap modes~\cite{Bornyakov:2007fm,Bornyakov:2008bg,Bornyakov:2008im},
we have turned to pure $SU(3)$ lattice gauge theory. The aim was to
find again hints for dyon structures (as topological clusters) very
close to the deconfinement temperature, revealed by the topological 
charge density defined with the massless overlap Dirac operator. 
An infrared scale is introduced by restriction
to a small number of modes of the overlap Dirac operator with low-lying
eigenvalues (``fermionic filtering''), i. e. only zero modes and near-zero
modes. In this analysis, three different types of boundary conditions have been
applied. The motivation was that these clusters might eventually be viewed
as dyons or antidyons, i.e. constituents of KvBLL calorons or anticalorons
\cite{Kraan:1998pm,Kraan:1998sn,Lee:1998bb} which come in three varieties.

We have demonstrated how their abundance and the tendency either to recombine
into calorons or to form pairs of different types depends on the temperature
in the vicinity of the deconfinement phase transition at $T_d \simeq 300$ MeV.
An increasing caloron dissociation has been observed when passing the
transition towards temperature values slightly above $T_d$.
Similar observations had been reported earlier for the $SU(2)$ Yang-Mills
case~\cite{Bornyakov:2007fm,Bornyakov:2008im}.

KvBLL (anti)calorons are (anti)selfdual solutions
of the classical $SU(3)$ Yang-Mills field equations with topological charge
$Q_t=\pm 1$ and $x_4$-periodicity (the latter related to the inverse
temperature). They are exhibiting very characteristic features worth to
look for in gauge field configurations provided by lattice simulations.
In case of the gauge group $SU(3)$ they consist of three constituents
(monopoles or ``dyons'') into which calorons can dissolve under specific
conditions, such that the dyon centers become separated sufficiently far
away from each other. The constituents in this limit become ``dyons''
well-separated and static in ``time'' $x_4$.
Their integrated action values or ``masses'' are fully determined by the
Polyakov loop at spatial infinity (``asymptotic holonomy''),
while their locations can be identified as positions, where the local
holonomy has at least two identical eigenvalues. Moreover, the zero-mode as
well as the low-lying eigenmodes of the massless Dirac operator become
localized around only one of the constituents \cite{Chernodub:1999wg}. 
On which constituent this happens, depends on the temporal boundary 
condition imposed on the Dirac operator defined in a finite space-time box.
(See the Appendix in \cite{Ilgenfritz:2013oda} which presents a brief summary
of those aspects of caloron solutions essential also for our present study.)

Suppose for a moment that these objects,
which are minimizing the Yang-Mills action, saturate the partition function
in a semiclassical-like path-integral representation. One is tempted to assume
that the ensemble-averaged Polyakov loop (as an order parameter for the
deconfinement transition) determines their asymptotic holonomy and therefore,
in particular, also the mass-ratio among the constituents of different types.
Since for $T \gg T_d$ the spatially averaged Polyakov loop tends to one of
the center values of $SU(3)$, one may expect that deeper in the deconfinement
phase one type of the tentative dyon constituents becomes very heavy, while
the others are light. Taking the statistical weight into account for the
constituents of different type, the heavy constituents should be suppressed
compared to the light ones. Then calorons as joint objects become more and
more suppressed, too, and we are left only with many light selfdual and
antiselfdual monopole constituents
which negligibly contribute to the topological charge. This might explain,
why the topological susceptibility becomes suppressed
at $T > T_d$ (in addition to the theoretically well-understood suppression of
caloron sizes \cite{Gross:1980br,Diakonov:2004jn}), while still an area
law of space-like Wilson loops is observed. This picture, which emerged from
our earlier $SU(2)$ lattice investigations \cite{Bornyakov:2007fm,Bornyakov:2008im},
has not yet been confirmed for the more realistic case of $SU(3)$ gluodynamics
at temperatures well above $T_d$.

In the lattice study of Ref. \cite{Ilgenfritz:2013oda} we considered only
$T$-values very close to $T_d$ where the spatial average of the trace of the
Polyakov loop was still fluctuating closely around the origin of the complex
plane. In order to confirm the appearance of topological objects like KvBLL
calorons and their dyon constituents we studied the low-lying spectrum of the
overlap Dirac operator together with the spatially local holonomy distribution.
The latter required an appropriate, small number of (over-improved) cooling
steps \cite{GarciaPerez:1993ki}. This cooling was found to shift the asymptotic
holonomy towards the respective $SU(3)$ center values and to influence also
the local holonomy inside topological clusters,
which primarily have been determined by the low-lying
overlap eigenmodes~\cite{Bornyakov:2013iva} of the un-cooled configurations.
The consequence of this exercise was that the positions of approximate equality
of two eigenvalues of the local holonomy became nicely correlated
with the centers of topological clusters. This gave us confidence that what
we are seeing at intermediate scales (of few lattice spacings),
can be interpreted as (anti)caloron and (anti)dyon excitations as described by
KvBLL solutions.

Here, in the present paper we are going a step further. At first, we are going
to higher temperature $~T \approx 1.5~T_d~$, and secondly, we employ another
feature which becomes important in the deconfined phase:  thermal  monopoles on
nearly static world lines
\cite{Bornyakov:1991se,Chernodub:2009hc,Chernodub:2006gu,Bornyakov:2013mxa}.  
Thermal monopoles are loops of magnetic currents
wrapped around the $x_4$ direction. As we shall see, they are also characterizing
dyon constituents by the occurrence of Abelian monopole world lines at their
centers. Therefore, we transform our real lattice gauge field into the maximally
Abelian gauge (MAG) and determine the corresponding magnetic currents after
Abelian projection. In order to clarify the role of cooling for the Abelian
monopole structure, we perform the gauge-fixing in two variants, without
(before) and with (after) cooling (that we again apply as few steps of
over-improved cooling). We shall convince ourselves that the thermal monopoles
are clearly correlated with the topological cluster centers determined from
fermionic filtering.  As a by-product, it turns out that  thermal monopoles
are rather stable under (moderate) cooling, in contrast to the local holonomy.

In \Sec{sec:lattice_setup} we introduce our lattice set-up, and
in \Sec{sec:topological_observables} we define the topologically
relevant lattice observables employed lateron.
In \Sec{sec:analytic_dyons}, starting from analytic KvBLL caloron solutions,
we construct (on a lattice) model gauge field configurations consisting of
one heavy or two light dyon-antidyon pairs.
We discuss how these pairs look like from the three points of
view that we shall apply in the following
also to analyze Monte Carlo generated thermal lattice gauge field
configurations of $SU(3)$ gluodynamics:
i) from the topological cluster analysis based on the low-lying
eigenmodes of the overlap operator,
ii) from the behavior of the local holonomy and its eigenvalues, and
iii) from the point of view of the MAG monopole current structure.
Light dyon-antidyon pairs are supposedly the prototype of topological
excitations in the bulk which guarantee the vanishing of the topological
susceptibility. Moreover, we expect a dilute-gas admixture of rare
and uncorrelated heavy (anti)dyon excitations.
Our model configurations will easily allow to distinguish
between light and heavy dyon-antidyon pairs.

Then, in \Sec{sec:ensemble_results} real gluodynamics is considered.
The occurence of a gap in the overlap eigenvalue spectrum depending
on the fermionic boundary condition is demonstrated and compared to
the reference cases of semi-analytical dyon-antidyon pairs.
We construct the fermionic topological charge density with the help
of a set of low-lying Dirac eigenmodes for all three different fermionic
boundary conditions. Finally, the clusters of the three densities under
consideration are presented and their correlation to the local holonomy
and to the static Abelian monopoles (obtained from the MAG construction)
is analyzed.

As a result, we shall clearly identify a large fraction of light (anti)dyons
and a smaller contribution of heavy (anti)dyons in the thermal lattice gluon
fields in agreement with the qualitative picture of the deconfinement phase
drawn above.

In \Sec{sec:conclusions} we shall draw our main conclusions.

\section{Lattice setup}
\label{sec:lattice_setup}

An ensemble of fifty $SU(3)$ gauge field configurations has been generated
for this investigation by sampling the pure $SU(3)$ gauge theory on a lattice
of size $20^3 \times 4$. We have used the L\"uscher-Weisz action
~\cite{Luscher:1984xn} with the value of the inverse coupling  $\beta = 8.25 $.
In our previous work~\cite{Ilgenfritz:2013oda} we were using the same action
at the same  $\beta= 8.25 $ but on a lattice of size $20^3 \times 6$.
This choice was meant to describe configurations slightly above the deconfining
temperature of $T_d \simeq 300$ MeV characteristic for pure $SU(3)$
gauge theory. This means that we are now addressing the deconfining phase
at a temperature $T \simeq 1.5~T_d$,
while the lattice discretization scale is about $~a \simeq 0.11$ fm.

Improved gauge actions are known to be mandatory for analyses using the overlap
Dirac operator, in order to take full advantage of the good chiral properties
of the latter. For example, the sampled gauge fields are smoother than
those sampled with the Wilson action. In particular, the idea of our analysis
rests on the observation that changing the boundary condition leaves the
number of zero modes unchanged, if the field is smooth enough.
The L\"uscher-Weisz action has also been used in the QCDSF topological
studies~\cite{Ilgenfritz:2007xu} of pure Yang-Mills theory with overlap
fermions and by Gattringer and
coworkers~\cite{Gattringer:2001jf,Gattringer:2001ia} when they were using a
specific chirally improved fermion action for topological investigations.
In the $SU(2)$ case, the tadpole-improved Symanzik action has been applied
for analogous reasons in our previous
work~\cite{Bornyakov:2007fm,Bornyakov:2008bg,Bornyakov:2008im}.

In addition to the plaquette term (pl), the L\"uscher-Weisz action includes
a sum over all $2 \! \times \! 1$  rectangles (rt) and a sum over all
parallelograms (pg), i.e.~all possible closed loops of length 6 along the
edges of all 3-cubes
\beqa
S[U]  = & \beta & \left(\sum_{pl} \frac{1}{3}  \mbox{Re~Tr}  [ 1 - U_{pl} ]\right.
\nonumber
\\
& + &  c_1 \sum_{rt} \frac{1}{3}  \mbox{Re~Tr}  [ 1 - U_{rt} ]
\\
& + & \left. c_2 \sum_{pg} \frac{1}{3}  \mbox{Re~Tr}  [ 1 - U_{pg} ] \right)\,,
\nonumber
\label{sgauge}
\eeqa
where the coefficients $c_1$ and $c_2$ are computed using results of one-loop
perturbation theory and tadpole
improvement~\cite{Luscher:1985zq,Snippe:1997ru,Lepage:1992xa}:
\beq
c_1 =  -  \frac{1}{ 20  u_0^2}
[ 1 + 0.4805  \alpha ]\,,~~
c_2 =  -  \frac{1}{u_0^2}  0.03325  \alpha \,.
\eeq
For the given $\beta=8.25$, the tadpole factor $u_0$  and the lattice
coupling constant $\alpha$ have been self-consistently determined
on a symmetric lattice ($20^4$) in a series of iterations via
\beq
u_0  =  \Big( \langle \frac{1}{3} \mbox{Re~Tr}~U_{pl} \rangle
\Big)^{1/4}\,, \quad \alpha  =  -
\frac{ \ln \Big( \langle \frac{1}{3} \mbox{Re~Tr}~U_{pl} \rangle
\Big)}{3.06839}
\eeq
arriving at the average plaquette value
$\langle \frac{1}{3} \mbox{Re~Tr}~U_{pl} \rangle = 0.639172$.

\section{Topologically relevant observables}
\label{sec:topological_observables}

The instruments (observables) of our analysis include
i) the local holonomy with its trace (i. e. the Polyakov loop),
ii) the (improved) gluonic topological charge,
iii) our definition of over-improved cooling,
iv) the Abelian monopoles revealed by Abelian projection in MAG,
and
v) the fermionic topological charge density and its ultraviolet-filtered
version, including the clustering properties of the latter.

In the context of topological structure, the meaning and usefulness of
the finite-temperature holonomy (considered globally and locally to
distinguish the dyonic constituents or ``instanton quarks'')
has become recognized only through the discovery of the KvBLL-caloron
solutions \cite{Kraan:1998pm,Kraan:1998sn,Lee:1998bb}.

\subsection{Holonomy}
\label{sec:holonomy}

Let us begin with the local holonomy and its eigenvalues.
The local holonomy is the product of timelike links
\beq
P(\vec{x}) = \prod_{x_0=1}^{N_\tau} U_0(\vec{x},x_0)
\label{eq:localholonomy}
\eeq
having eigenvalues
\beq
\lambda_k=\exp\left(i 2\pi \mu_k(\vec{x})\right)
\label{eq:localholonomy_evs}
\eeq
which obviously depend on the spatial position.

On one hand, the spatial positions of the dyon constituents
of KvBLL caloron solutions are determined by the condition,
that two of these eigenvalues should coincide (cf. Appendix
in \cite{Ilgenfritz:2013oda}). Lateron we shall use this
property to localize (anti)dyons in artificially modelled
as well as in simulated lattice field configurations.
On the other hand, the asymptotic holonomy of KvBLL calorons
(after a suitable constant gauge transformation)
\beq
\pl \equiv \lim_{|\vec x|\rightarrow\infty} P(\vec{x}) =
\exp[2\pi i\,{\rm diag}(\mu_1,\mu_2,\mu_3)],
\eeq
with real and ordered numbers
$\mu_1\leq\ldots\leq\mu_3\leq\mu_{4}\!\equiv\!1\!+\!\mu_1$ and
$~\mu_1 + \mu_2 + \mu_3 \!=\!0$)
determines the masses of well-separated dyon constituents via
$8\pi^2\nu_m$, where $\nu_m\!\equiv\!\mu_{m+1}\!-\!\mu_m$
(cf. Appendix in \cite{Ilgenfritz:2013oda}).

Taking the trace of $P(\vec{x})$ one gets the (gauge invariant)
complex-valued Polyakov loop
\beq
L(\vec{x}) = \frac{1}{3} \Tr~P(\vec{x}) \; .
\eeq
For $SU(3)$, the condition of two coinciding eigenvalues of the
local holonomy corresponds to the respective Polyakov loop being
located on the periphery of the Polyakov triangle. All three
eigenvalues coincide only in its corners with
$L(\vec{x})=z_k=\exp\left(i~k~2~\pi/3\right) \cdot {\bf 1}$
with $k=0,1,2$.
We call $\overline L$ the spatially averaged Polyakov loop of a
given gauge field configuration:
\beq
\overline L = \frac{1}{V}~\sum_{\vec{x}} L(\vec{x}) \; .
\eeq

Averaged appropriately over the statistical ensemble of gauge fields
it serves as an order parameter for the deconfinement transition
of pure Yang-Mills theory.
The latter has a global $Z(3)$ symmetry of the action, and the
deconfined phase is characterized by the spontaneous breaking of this
symmetry, i. e.  the non-vanishing spatially averaged Polyakov loop,
falling into one of the $3$ sectors, can be represented as
$\overline L  \approx |\overline L| z_k.$
For finite systems, transitions between the $Z(3)$ sectors (i. e.
between different deconfined phases) are not excluded. Therefore,
we consider all configurations with non-vanishing $\overline{L}$
transformed by a $Z(3)$ flip to the real sector where
$\overline L  \approx |\overline L| z_0$.

\subsection{The gluonic definition of the topological density}
\label{sec:topdensity}

The continuum definition of topological charge density is
\begin{equation}
q(x) =  \frac{1}{16 \pi^2} \Tr (F_{\mu\nu}(x)\,\tilde{F}_{\mu\nu}(x))
\label{eq:qdensity}
\end{equation}
with
\begin{equation}
\tilde{F}_{\mu\nu}(x)=
 \frac{1}{2}~\epsilon_{\mu\nu\lambda\sigma}~F_{\lambda\sigma}(x)\,.
\end{equation}
The (improved) gluonic topological charge density on the lattice
rests on the field strength definition of $F^{(n)}_{\mu\nu}(x)$
as a ``clover'' average over all untraced plaquette loops (with sidelength
$n=1$) and over four untraced extended, quadratic loops of size
$n \times n$ (with $n=2,3$ in the improved case) within the $\mu\nu$
plane, placed around each site $x$ and kept untraced in that site
$x$ ~\cite{BilsonThompson:2002jk}. The improved topological charge and
the corresponding continuum action (in units of the one-instanton action
$S_{\rm inst}$) are then defined as
\begin{eqnarray}
Q_{\rm glue} &=& \sum_x q(x) \label{gluonic_Q}\,, \\
S/ S_{\rm inst} & = & \sum_x \Tr (F_{\mu\nu} (x)~F_{\mu\nu} (x)) / (16 \pi^2)\,.
\label{gluonic_S}
\end{eqnarray}

\subsection{Cooling and over-improved cooling}
\label{sec:cooling}

Cooling is an {\it ad hoc} method to remove quantum fluctuations
up to a certain ``diffusion'' scale from given lattice field
configurations created in the course of Monte Carlo simulations.
It proceeds by a sweep over the lattice links, where
the link is updated in such a way to warrant the minimum of action
relative only to this link $U_{x,\mu} \in SU(3)$ while all other links
remain untouched. For the $SU(3)$ gauge group this local minimization
is realized in form of a sweep over three $SU(2)$ subgroups of $SU(3)$
(Cabibbo-Marinari method). Cooling can be defined with respect to
different gluonic actions, not necessarily the action used for the Monte
Carlo generation of the ensemble to work on. The simplest case is with
respect to the Wilson (one-plaquette) action. More generally, cooling
is defined with respect of an action of the form
\begin{eqnarray}
S(\epsilon) &=& \sum_{x,\mu\nu}  \frac{4-\epsilon}{3}
            \mbox{Re~Tr}~\left( 1 - U_{x,\mu\nu} \right) \nonumber \\
            &+& \sum_{x,\mu\nu}  \frac{1-\epsilon}{48}
            \mbox{Re~Tr}~\left( 1 - U^{2\times2}_{x,\mu\nu} \right)
\end{eqnarray}
which reduces to Wilson action in the case $\epsilon=1$. The so-called
over-improved action
\cite{GarciaPerez:1993ki}
corresponds to $\epsilon=-1$. Expanding in powers of lattice spacing $a$
one finds
\begin{equation}
S(\epsilon) = \sum_{x,\mu\nu} a^4  \Tr \left[  \frac{1}{2} F^2_{\mu \nu}(x)
- \frac{\epsilon a^2}{12} \left(D_\mu F_{\mu \nu}(x) \right)^2 \right]+O(a^8).
\end{equation}
For a discretized continuum instanton of size $\rho$ this provides
\begin{equation}
S(\epsilon) = 8 \pi^2 \left[ 1 - \frac{\epsilon}{5} \left(\frac{a}{\rho}\right)^2
+ {\cal O}\left(\left[\frac{a}{\rho}\right]^4\right) \right]
\end{equation}
suggesting that $\rho$ under cooling will decrease for $\epsilon > 0$
and increase for $\epsilon < 0$.
The inversion of lattice artefacts relative to the Wilson case makes
topological lumps stable against cooling.

It is worth noting that standard cooling ($\epsilon=1$) - averaged over
gauge field ensembles - can be nicely mapped one-to-one \cite{Bonati:2014tqa}
to the theoretically well-understood Wilson or gradient flow
\cite{Luscher:2009eq,Luscher:2010iy,Luscher:2011bx}. We believe the
same to hold for over-improved cooling.

\subsection{The overlap Dirac operator, the near-zero band and the UV filtered
topological density}
\label{sec:overlap}

The next of our tools is the construction of the near-zero-mode
band of eigenmodes of the massless overlap operator $D$. The overlap
Dirac operator $D$  fulfills the Ginsparg-Wilson
equation~\cite{Ginsparg:1981bj}. A possible solution -- for any
{\it input Dirac operator}, in our case for the Wilson-Dirac operator
$D_W$ -- is the following zero-mass overlap Dirac
operator~\cite{Neuberger:1997fp,Neuberger:1998wv}
\begin{eqnarray}
D(m=0)&=&\frac{\rho}{a}\,\left( 1 + \frac{D_W}{\sqrt{D_W^{\dagger}\,D_W}}
\right) \nonumber \\
&=&\frac{\rho}{a}\,\left( 1 + {\rm sgn}(D_W) \right) \,,
\label{eq:OverlapDirac}
\end{eqnarray}
with $D_W = M - \frac{\rho}{a}$, where $M$ is the hopping term of the
Wilson-Dirac operator and $\frac{\rho}{a}$ a negative mass term usually
subject to optimization.
The index of $D$, i.e. the difference of its number of right-handed and left-handed
zero modes, can be identified with the integer-valued topological charge
$Q_{\rm over}$ \cite{Hasenfratz:1998ri}.

The topological charge density with maximal resolution (down to the lattice
spacing $a$) is defined in terms of the overlap Dirac operator as
follows
\begin{equation}
q(x) = - {\rm tr} \left[ \gamma_5 \left( 1
       - \frac{a}{2}\,D(m=0;x,x) \right)\, \right] \,.
\label{eq:TopDensI}
\end{equation}
Using the spectral representation of (\ref{eq:TopDensI}) after
diagonalization (using a variant of the Arnoldi algorithm)
in terms of the eigenmodes $\psi_{\lambda}(x)$ with
eigenvalue $\lambda$, an UV smoothed form of the density can be defined
by filtering, i. e. summing over a narrow band of near-zero eigenmodes:
\begin{equation}
q_{\lambda_{\rm sm}}(x) = - \sum_{|\lambda| < \lambda_{\rm sm}}
\left( 1 - \frac{\lambda}{2} \right)
\,\sum_c \left( \psi_{\lambda}^c(x)\, ,\gamma_5\, \psi_{\lambda}^c(x) \right) 
\label{eq:TopDensII}
\end{equation}
summed over color $c$ and with $\lambda_{\rm sm}$ as an UV cutoff.

While the physical fermion sea is described by a Dirac operator with
antiperiodic boundary condition, for the analysis of topological structure
it is useful to diagonalize the overlap Dirac operator with a continuously
modified boundary condition, which is characterized by an angle $\phi$,
\beq
\psi(\vec{x},x_4+\beta) = \exp(i\phi)\psi(\vec{x},x_4) \; .
\label{eq:bc1}
\eeq
We have chosen three angles
\beq
\phi = \left\{
\begin{array}{ll}
 \phi_1  \equiv -\pi/3\, \\
 \phi_2  \equiv +\pi/3\, \\
 \phi_3  \equiv ~~~~\pi\,
\end{array}
\right\}
\label{eq:bc2}
\eeq
\noindent
ensuring for a single caloron solution the corresponding fermion zero modes
to become maximally localized at each, but always at one of its three
constituent dyons. Note that $\phi_3$ corresponds to the antiperiodic
boundary condition. Only the spectrum corresponding to the latter boundary
condition (no matter whether the gauge field ensemble is quenched or not)
develops a gap in the high temperature phase (see Ref. \cite{Bilgici:2009tx}).

On the contrary, for the confined phase of gluodynamics (as well as
for the chirally broken phase of QCD) it is known that the gross features
of the Dirac spectrum are only weakly dependent on the boundary conditions.

In the deconfined (high temperature) phase, the construction of the UV
smoothed topological charge density
in terms of the eigenvalues and eigenmodes
should be specifically done for the three boundary conditions:
\beq
q_{i,N}(x) = - \sum_{j=1}^N
\left( 1 - \frac{\lambda_{i,j}}{2} \right)
\,\sum_c \left( \psi_{i,j}^c(x)\, ,\gamma_5\, \psi_{i,j}^c(x) \right) \, ,
\label{eq:truncated_density}
\eeq
where $ i =1,2,3$ enumerates the three boundary conditions defined by \Eq{eq:bc2}
and $j$ enumerates the eigenvalues $\lambda_{i,j}$ arranged in increasing order
$ \lambda_{i,1} <  \lambda_{i,2} < ... < \lambda_{i,N} $.

Let us note here, that topologically non-trivial clusters filtered out
with the truncated densities (\ref{eq:truncated_density}) and averaged over the
boundary conditions can be nicely mapped onto the ones seen with the
gluonic definition (\ref{eq:qdensity}) after an optimized number of
APE or STOUT smearing steps
\cite{Bruckmann:2006wf,Ilgenfritz:2008ia,Bruckmann:2009vb}. We have no
doubt that this will hold also when applying the Wilson flow
accordingly \cite{Luscher:2009eq,Luscher:2010iy,Luscher:2011bx}.

\subsection{MAG and Abelian monopoles definitions}
\label{sec:mag-monopolesa}
We use the definition of MAG introduced for
lattice $SU(N)$ theory in \cite{Kronfeld:1987vd} and later specified for
the $SU(3)$ group in \cite{Brandstater:1991sn}.
The MAG is fixed by maximizing the functional
\beqa
\label{FU}
F[U] = \frac{1}{12\,V}~
 \sum_{x,\mu}~\left[ |(U_\mu(x))_{11}|^2 +|(U_\mu(x))_{22}|^2 \right. \nonumber \\
            \left. +|(U_\mu(x))_{33}|^2 \right]
\eeqa
with respect to local gauge transformations $g$ of the lattice gauge field,
\beq
U_\mu(x) \to U^g_\mu(x) = g(x)^\dagger U_\mu(x) g(x+\hat \mu)\,.
\eeq
Note that alternative definitions of the MAG for the $SU(3)$ group were
introduced in \cite{Tucker:2001tt} and were recently studied in
\cite{Bonati:2013bga}.
We use the simulated annealing algorithm first used to fix MAG in the
$SU(2)$ case \cite{Bali:1996dm} and then extended to the $SU(3)$ group in
\cite{Bornyakov:2001qw}. The details of the implementation of simulated
annealing for the case of the $SU(3)$ gauge group can be found in
\cite{Bornyakov:2003vx}. 
To reduce the effects of ambiguities due to
Gribov copies we have always generated 10 random gauge copies and have
picked up the copy with the maximal value of the gauge fixing functional.

The Abelian field $u_\mu(x) \in U(1) \times U(1)$ is determined as
\beq
\label{uabel}
u_\mu(x) = \mbox{diag} \left(u^{(1)}_\mu(x),u^{(2)}_\mu(x),u^{(3)}_\mu(x)
\right)\,,
\eeq
where
\beq
\label{ulink}
u^{(a)}_{\mu}(x)=e^{i \theta^{(a)}_{\mu}(x)} 
\eeq
with
\beq
\label{tlink}
\theta^{(a)}_{\mu}(x) = \arg~(U_\mu(x))_aa-\frac{1}{3} \sum_{b=1}^3
\arg(U_\mu(x))_bb\,\big|_{\,{\rm mod}\ 2\pi}
\eeq
such that
\beq
\theta^{(a)}_{\mu}(x) \in [-\frac{4}{3}\pi, \frac{4}{3}\pi]\,.
\eeq
This definition of Abelian projection $u_\mu(x)$ maximizes the expression
$|\mbox{Tr} \left( U_\mu^\dagger(x) u_\mu(x) \right) |^2$
\cite{Bornyakov:2004ii}.

The monopole currents are residing on links of the dual lattice
and are defined by
\beq\label{current}
j^{(a)}_\mu(^*x) = \frac{1}{2\pi} \epsilon_{\mu\nu\alpha\beta}
\partial_\nu~\bar{\theta}^a_{\alpha\beta}(x) = 0,\pm 1,\pm 2 \, ,
\eeq
where $\partial_\nu$ is the forward lattice derivative,
and Abelian flux $\bar{\theta}^a_{\mu\nu} \in (-\pi, \pi] $ is defined
from the Abelian pla\-quette
\beq
\theta^{(a)}_{\mu\nu}(x) =
  \partial_\mu \theta^{(a)}_\nu (x) - \partial_\nu \theta^{(a)}_\mu (x)
\eeq
using the relation
\beq
\theta^{(a)}_{\mu\nu}(x) = \bar{\theta}^{(a)}_{\mu\nu}(x) + 2 \pi m^a_{\mu\nu}(x)\,.
\eeq
They are then shifted by $2\pi n$ to satisfy
\beq \label{steq0}
\sum_{a=1}^3 \bar{\theta}^{(a)}_{\mu\nu}(x) = 0 \, .
\eeq
This guarantees that
\beq
\sum_{a=1}^3 j^{(a)}_\mu(x) = 0 \, ,
\eeq
i.e. only two currents are independent.
The current conservation law is satisfied for every $a$ separately:
\beq
\sum_{\mu}\partial^-_\mu j^{(a)}_\mu(s) = 0 \,, \,\, a=1,2,3 \,.
\eeq

\section{Constructing (anti)dyon pairs}
\label{sec:analytic_dyons}

The analytic construction of an $SU(3)$ caloron is described in
Ref.~\cite{Kraan:1998sn}. It can be used in order to create model
configurations of the type we might expect to be dominant in the
thermalized gauge field configurations at $T=1.5~T_d$ studied
in this paper.

Therefore, we have chosen model calorons with an asymptotic holonomy
$\pl =\exp[2\pi i\,{\rm diag}(\mu_1,\mu_2,\mu_3)]$ with
$\mu_1= - \mu_3 = - 0.271$ and $\mu_2 = 0$
such that its Polyakov loop value is
$ L = \frac{1}{3} \Tr~\pl = (2 \cos(2 \pi \mu_1)+1)/3 = 0.24$.
This value corresponds to the averaged Polyakov loop
$\langle |\overline{L}| \rangle$ observed for the thermalized
gauge field configurations at $1.5~T_d$.

Then two of the three constituent dyons are light (i.e. if well-separated
they carry the topological charge fraction $\nu_1=\nu_2=0.271$) and the
third becomes heavier ($\nu_3=0.458$). The local Polyakov loops at the
dyon positions happen to be located on the sides of Polyakov triangle
as pointed out in the Appendix of Ref.~\cite{Ilgenfritz:2013oda}.

\begin{figure*}[!htb]
\centering
\includegraphics[width=.27\textwidth]{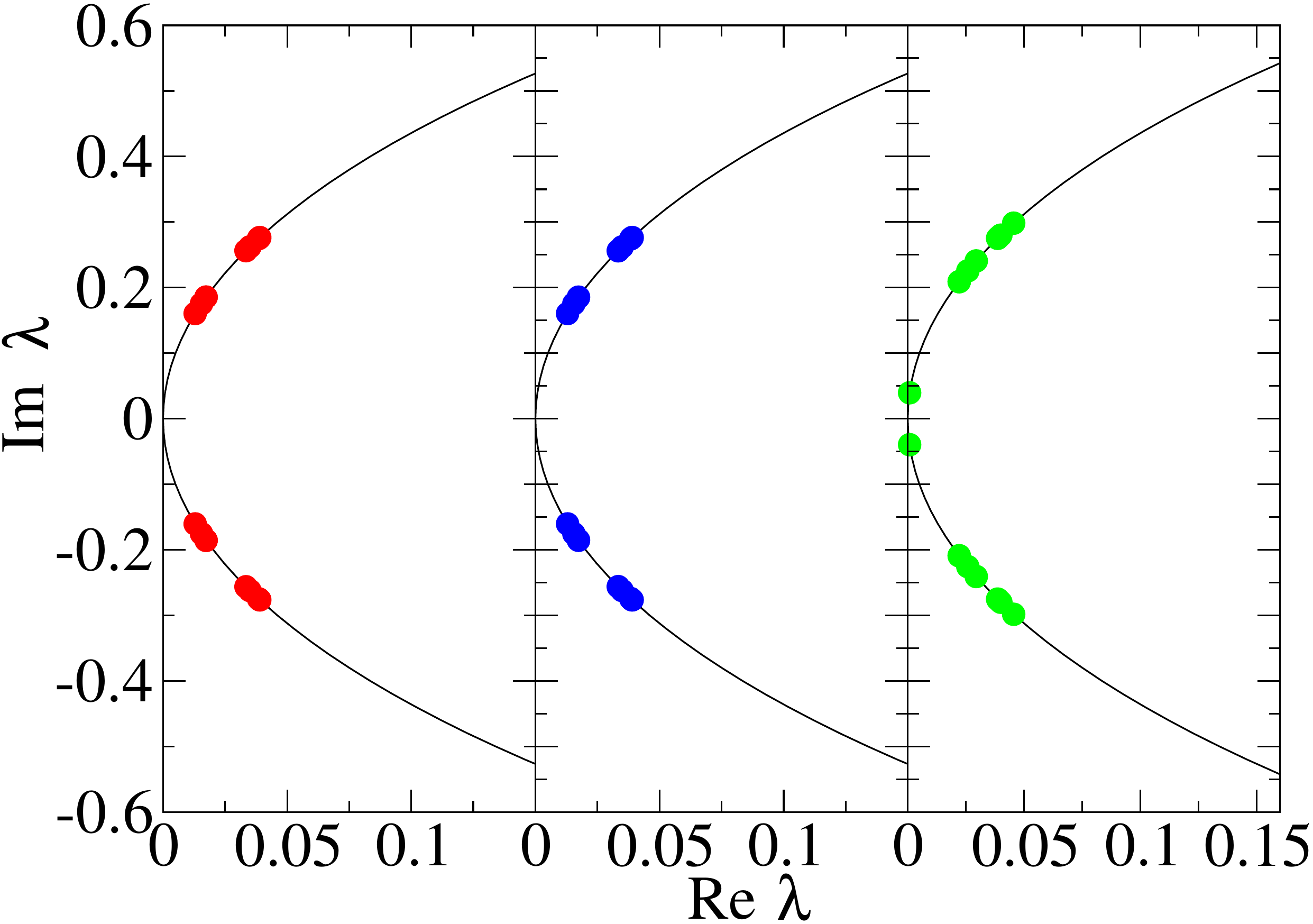}
\vspace{1cm}
\caption{Overlap eigenvalues for a heavy (third-type) dyon-antidyon pair for
the three different boundary conditions of \Eq{eq:bc2}.}
\label{spectad}
\end{figure*}
\begin{figure*}[!htb]
\centering
a) \includegraphics[width=.31\textwidth]{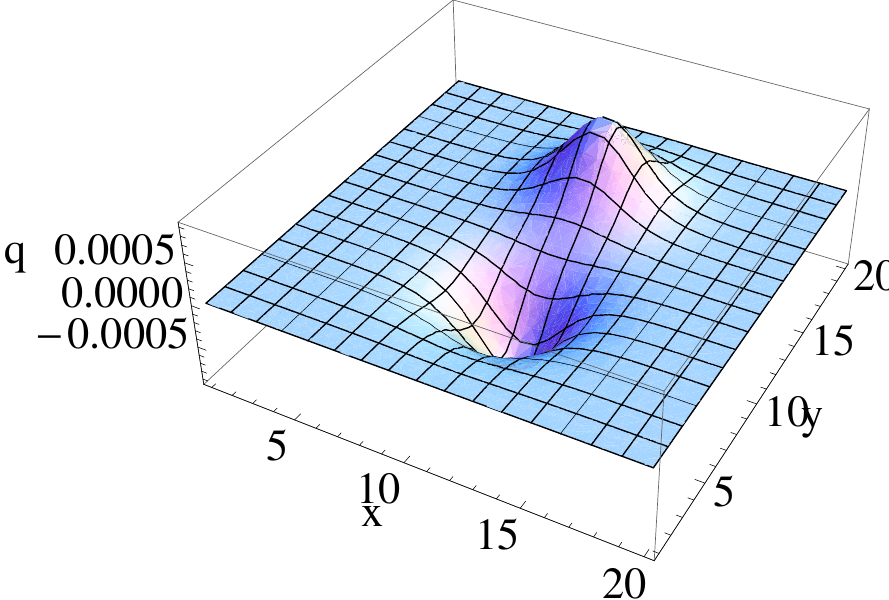}%
b) \includegraphics[width=.31\textwidth]{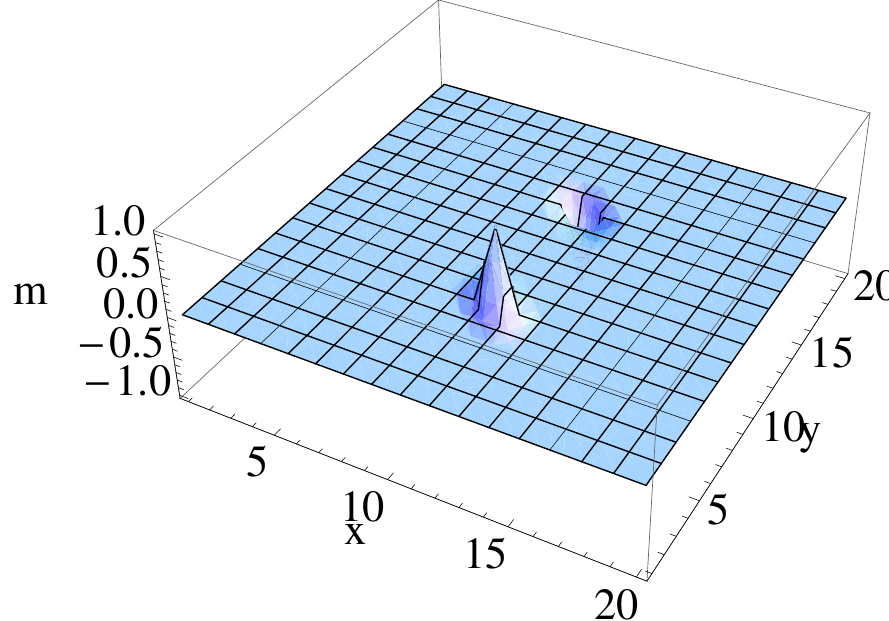}%
c) \hspace{0.1cm}\includegraphics[width=.27\textwidth]{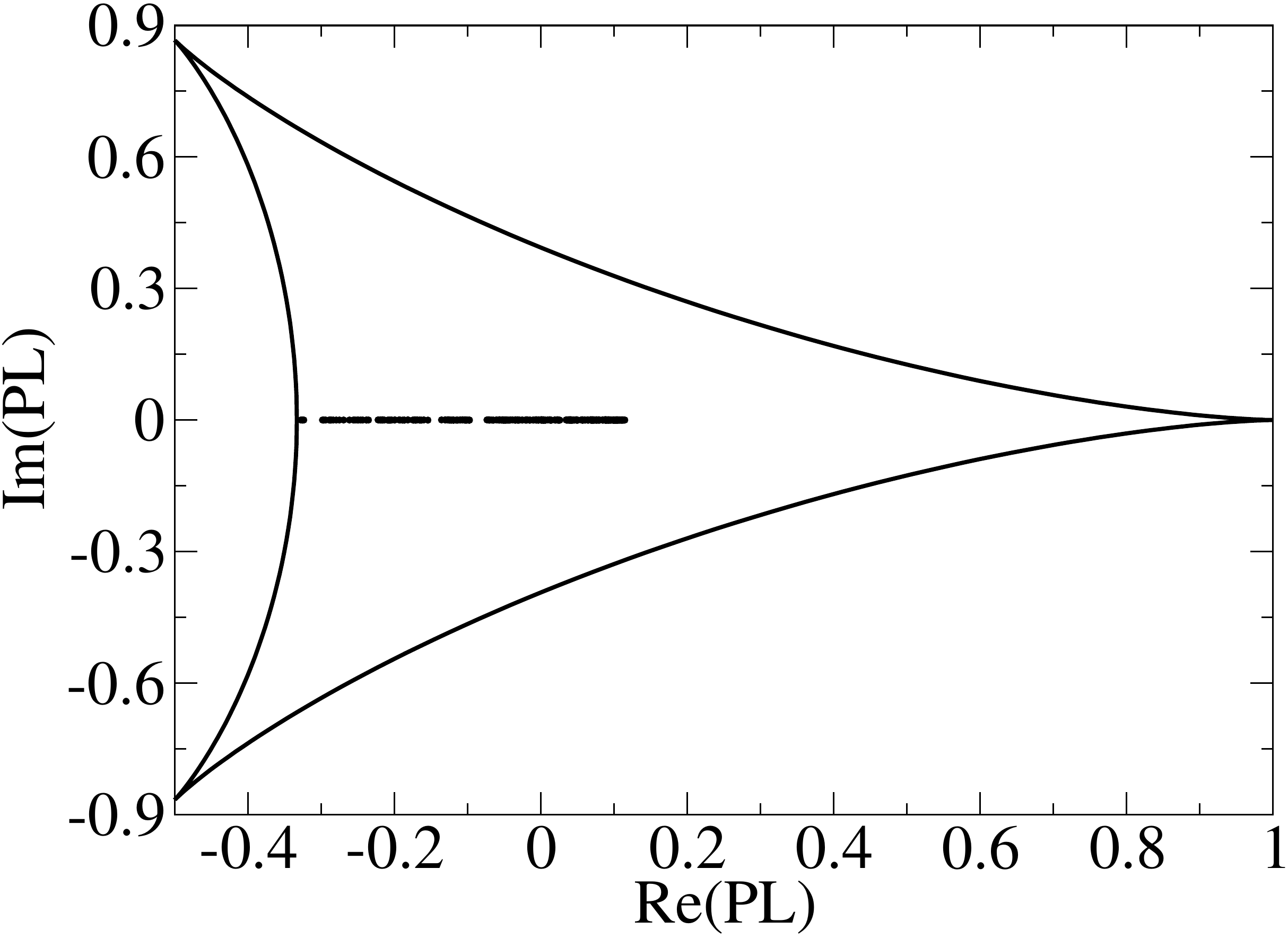}\\
\vspace{1cm}
\caption{A heavy (third-type) dyon-antidyon pair is presented
a) by the profile of gluonic topological charge density over the $xy$-plane,
b) by the local magnetic charge distribution
   of (static, timelike) MAG monopole currents in the $xy$-plane and
c) a scatter plot of the Polyakov loop values taken at
   the same plane.}
\label{tad}
\end{figure*}
\begin{figure*}[!htb]
\centering
a) \includegraphics[width=.31\textwidth]{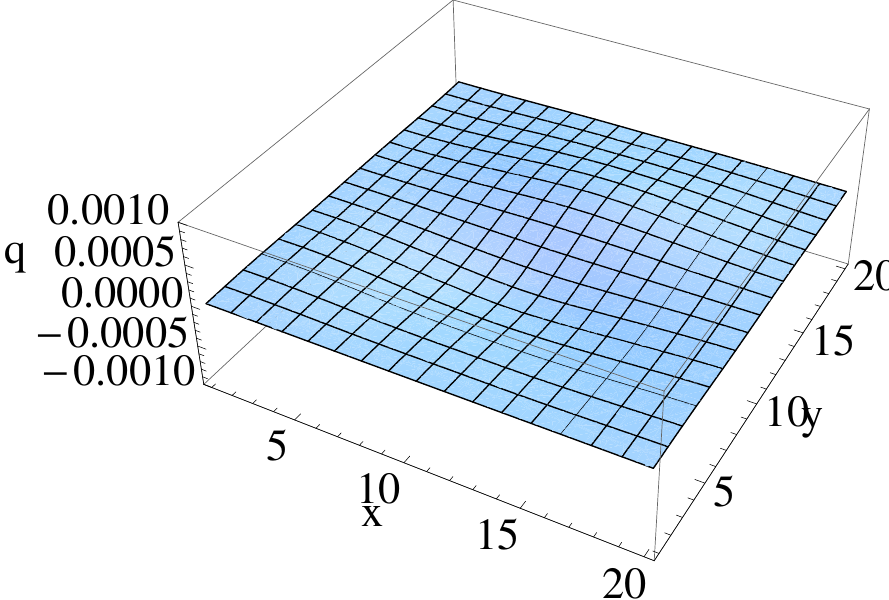}%
b) \includegraphics[width=.31\textwidth]{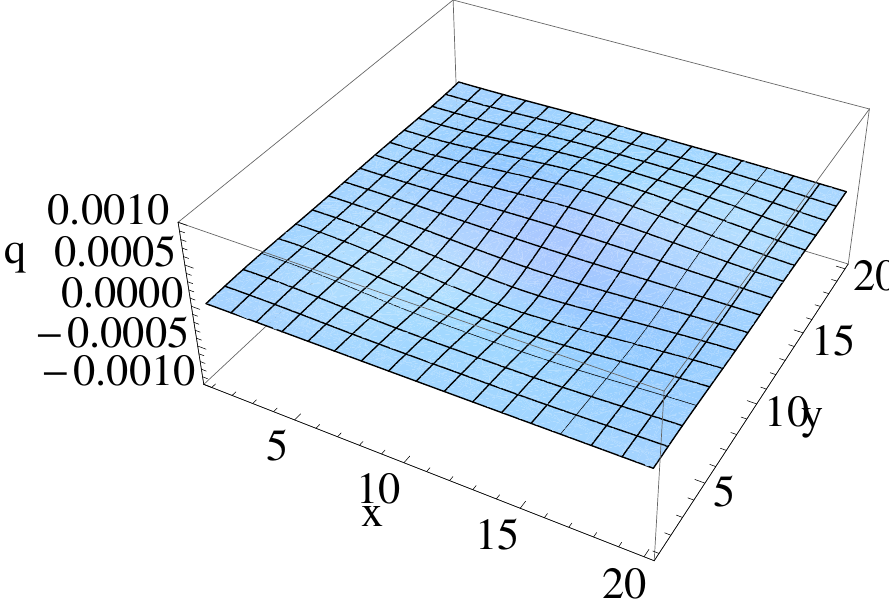}%
c) \includegraphics[width=.31\textwidth]{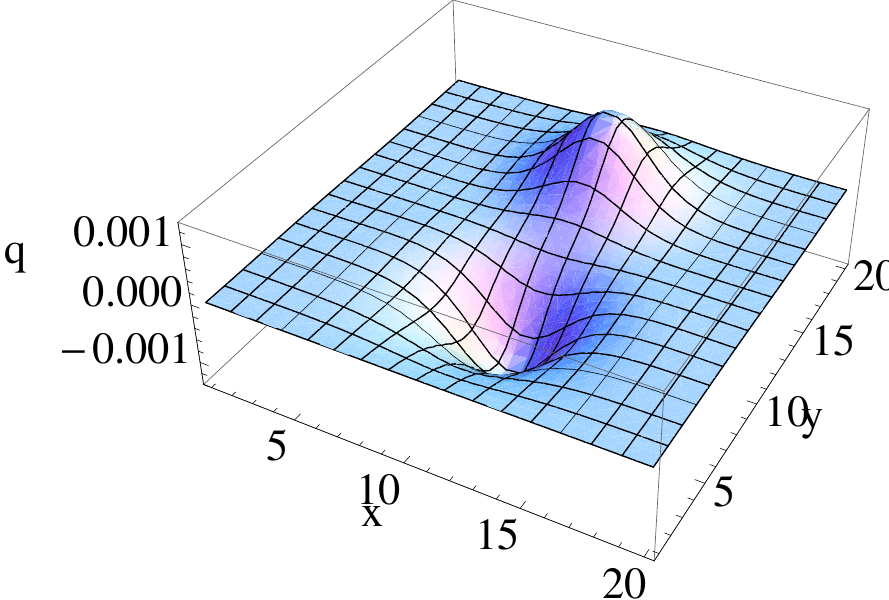}\\
\vspace{1cm}
\caption{The fermionic topological charge densities a),b),c) are shown for
the heavy dyon-antidyon pair reconstructed from the eigenmodes, corresponding
to the three types of boundary conditions.}
\label{overtad}
\end{figure*}

In order to construct a dyon-antidyon pair of the third (heavy) type
we have placed the dyon-triplet of such a caloron solution at positions
\beqa
 x_1=-1,  & ~~~ y_1=-10,  & ~~~ z_1=0\nonumber \\
 x_2=~1,  & ~~~ y_2=-10,  & ~~~ z_2=0\\
 x_3=~0,  & ~~~ y_3=~~1,  & ~~~ z_3=0\nonumber
\eeqa
with the local Polyakov loop values
$$ L(\vec{x}_1) = L(\vec{x}_2)^*= 0.395 - i*0.171 ,
~~L(\vec{x}_3) = -0.333 $$
and calculated the potentials for $-2.5<x,z<2.5$ in the positive-$y$
half-space $0<y<2.5$ (all numbers are given in units of the time period
of periodic caloron solution).
Note, that the light dyons are far outside the region where the gauge field
is calculated. Similarly, we have placed antidyons (as constituents of a
corresponding anticaloron solution) at
\beqa
 \bar x_1=-1,  & ~~~ \bar y_1=10,  & ~~~ \bar z_1=0\nonumber \\
 \bar x_2=~1,  & ~~~ \bar y_2=10,  & ~~~ \bar z_2=0\\
 \bar x_3=~0,  & ~~~ \bar y_3=-1,  & ~~~ \bar z_3=0\nonumber
\eeqa
and calculated their potentials for $-2.5<x,z<2.5$ in the negative-$y$
half-space $-2.5<y<0$.
Finally, we have sewed together these potentials defined in their
respective domains and - for the purpose of discretization -
have calculated links on a $20\times 20 \times 20\times 4$ lattice
(with lattice spacing equal to $1/4$). This lattice covers the full
spatial region $-2.5<x,y,z<2.5$ and the temporal periodicity range $0<t<1$.
On this lattice the dyon is at position $(10, 14, 10)$ and the antidyon at
$(10, 6, 10)$. Next, cooling has been applied to the constructed lattice
field configuration in order to remove the discontinuites left over from
sewing together the half-spaces and in order to arrange for
smooth spatially periodic boundary conditions (spatial torus).

We have analyzed the obtained configuration by diagonalizing the overlap
Dirac operator and identifying $N = 20$ near-zero eigenvalues and
respective eigenmodes. The diagonalization has been performed for three
temporal boundary conditions (\ref{eq:bc2}).
The pattern of near-zero eigenvalues of the overlap operator is shown
in \Fig{spectad} for the three types of boundary conditions.
Note that the spectrum develops a gap for the first and second
kind of boundary conditions, while there are near-zero
eigenvalues for the third kind.

The profile of the gluonic topological charge density (as described
in \Sec{sec:topdensity}) over the $xy$-plane that contains the dyon-antidyon
pair is shown in \Fig{tad}a. In the same plane static MAG  thermal
monopoles are found. Their positions are visualized in \Fig{tad}b.
Here and in the following we will have only temporal ($\mu=4$)
magnetic currents $j^{(a)}_4(x)$, and these of three types: $(\pm 1,\mp 1,0)$,
$(0,\pm 1,\mp 1)$, $(\mp 1,0,\pm 1)$, that will form closed loops in the
temporal direction.

For the given configuration, the same $xy$-plane is mapped to the complex
plane of the Polyakov loop. The scatter plot \Fig{tad}c shows the local
Polyakov loop of the isolated heavy dyon-antidyon pair as the part of the
real axis connecting the origin with the left side of the Polyakov triangle.

Using $N=20$ low-lying eigenmodes of the overlap Dirac operator
we have reconstructed the profiles of the fermionic
topological charge density according to the spectral representation of the
latter (\ref{eq:truncated_density}) for the three temporal boundary
conditions (\ref{eq:bc2}). Only the third boundary condition
catches the topological charge profile of the dyon-antidyon pair of third
(heavy) type (see \Fig{overtad}).

This example of a dyon-antidyon pair demonstrates that there is a strong
correlation between clusters of gluonic as well as fermionic topological
charge density on one hand and MAG monopoles on the other.

\begin{figure*}[!htb]
\centering
\includegraphics[width=.27\textwidth]{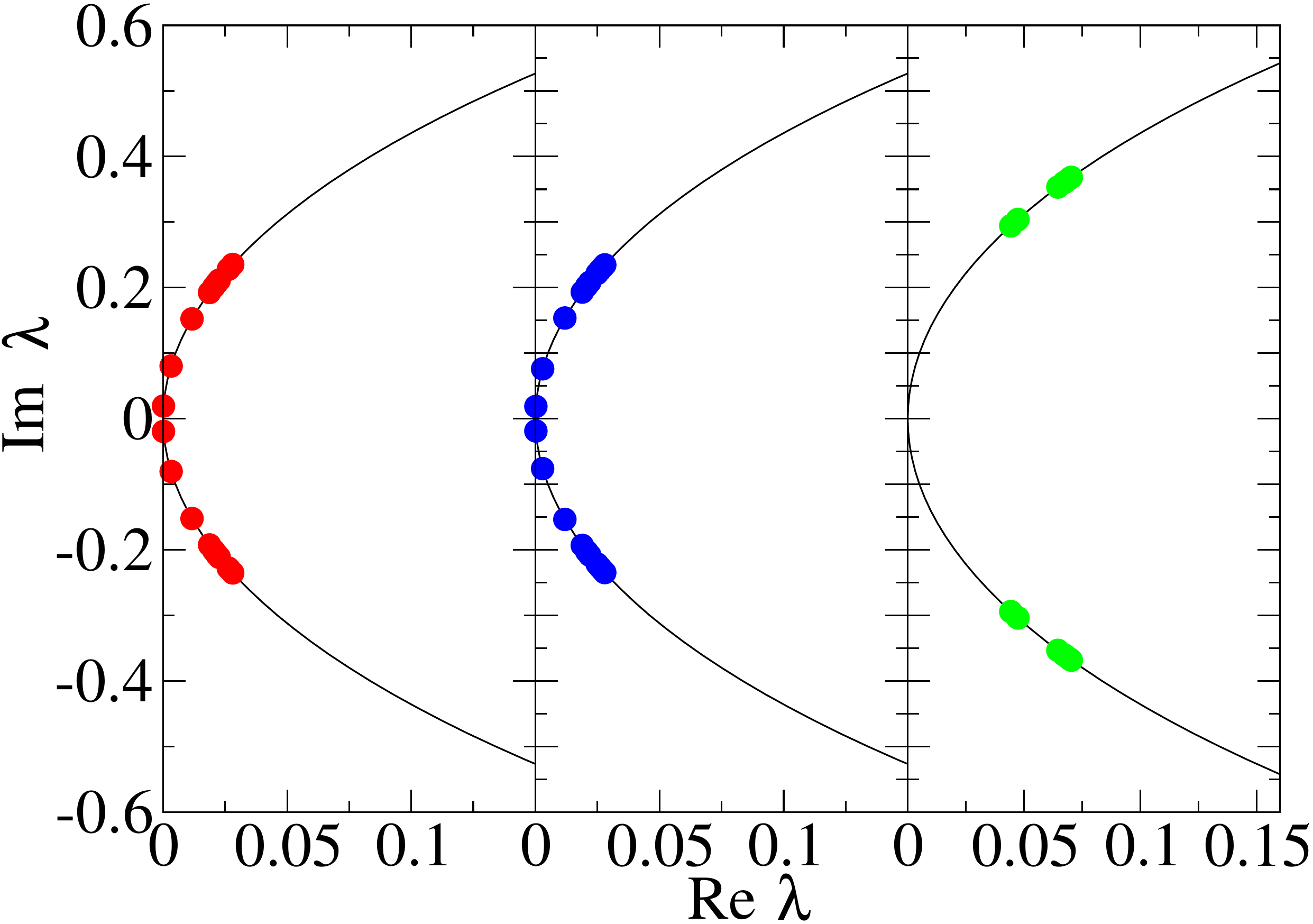}
\caption{Overlap eigenvalues for a light double-dyon-antidyon pair for the
three different boundary conditions of \Eq{eq:bc2}.}
\vspace{0.5cm}
\label{specdad}
\end{figure*}
\begin{figure*}[!htb]
\centering
a)\includegraphics[width=.31\textwidth]{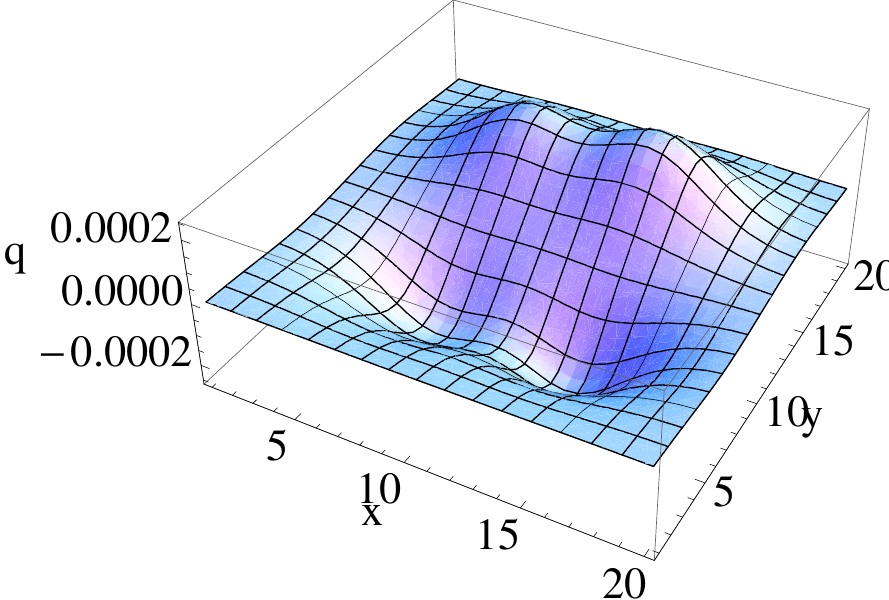}%
b) \includegraphics[width=.31\textwidth]{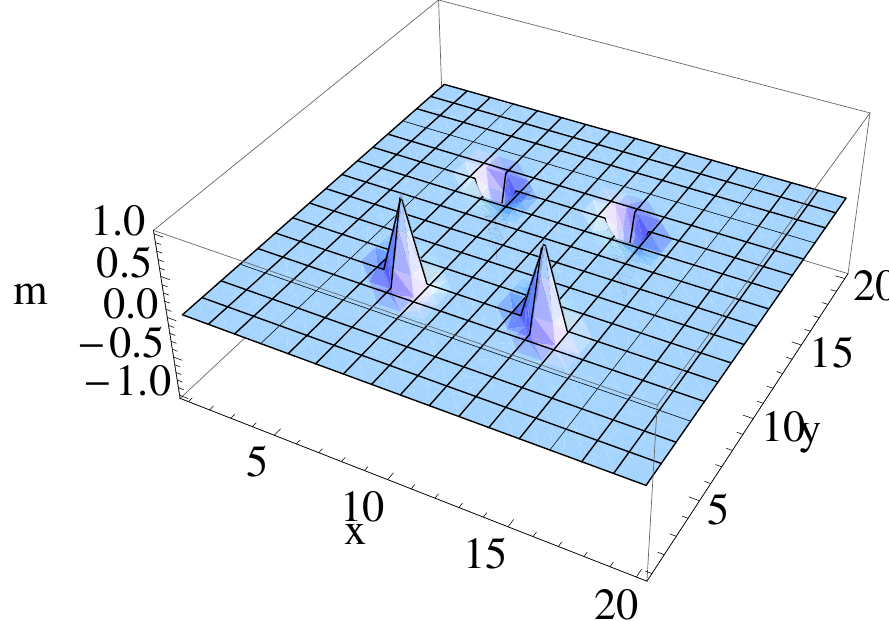}%
c)\hspace{0.1cm}\includegraphics[width=.27\textwidth]{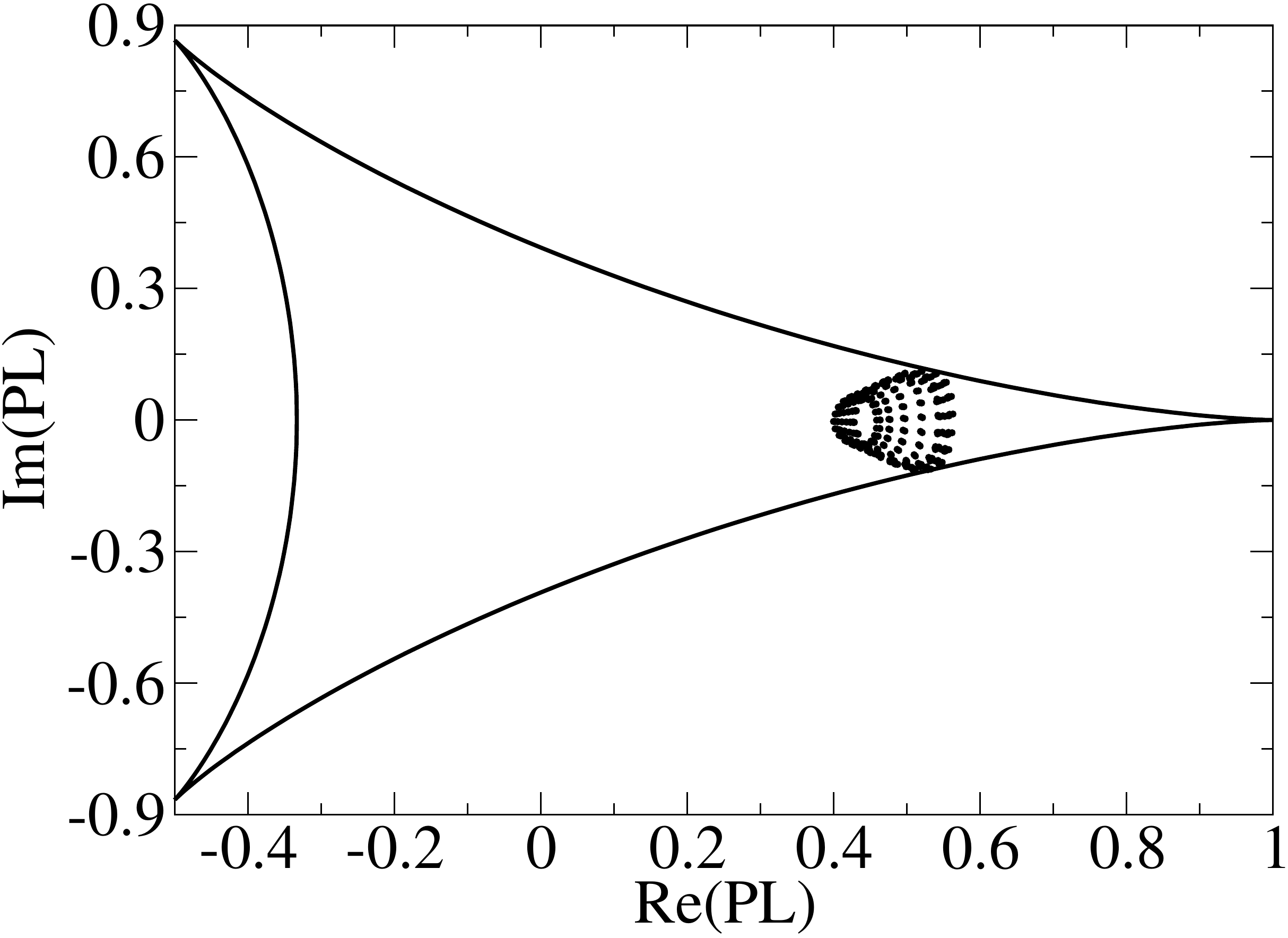}\\
\caption{For the light double-dyon-antidyon pair we show
a) the profile of gluonic topological charge density on the $xy$-plane,
b) the local magnetic charge distribution of (static, timelike) MAG
monopole currents in the $xy$-plane and
c) a scatter plot of the Polyakov loop values picked up at the same plane.}
\label{dad}
\end{figure*}
\begin{figure*}[!htb]
\centering
 a) \includegraphics[width=.31\textwidth]{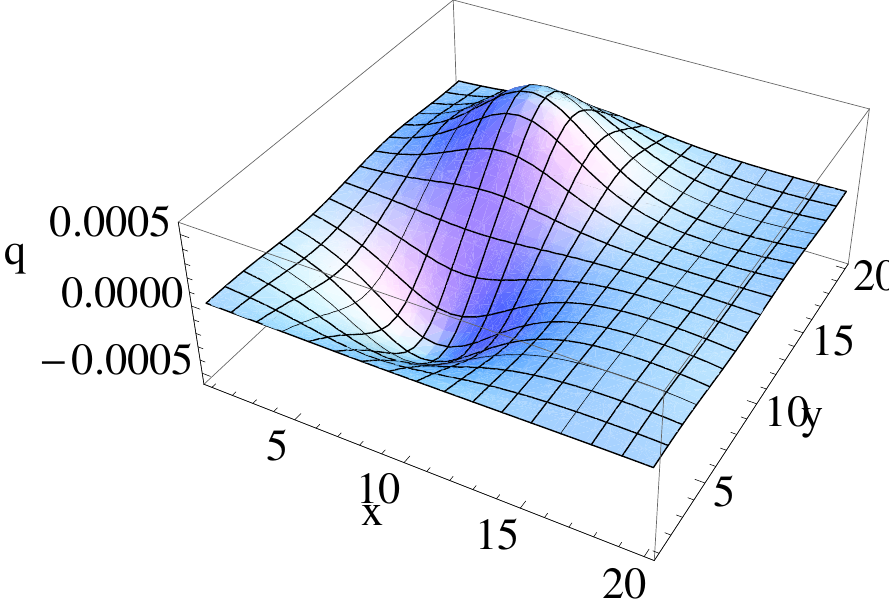}%
 b) \includegraphics[width=.31\textwidth]{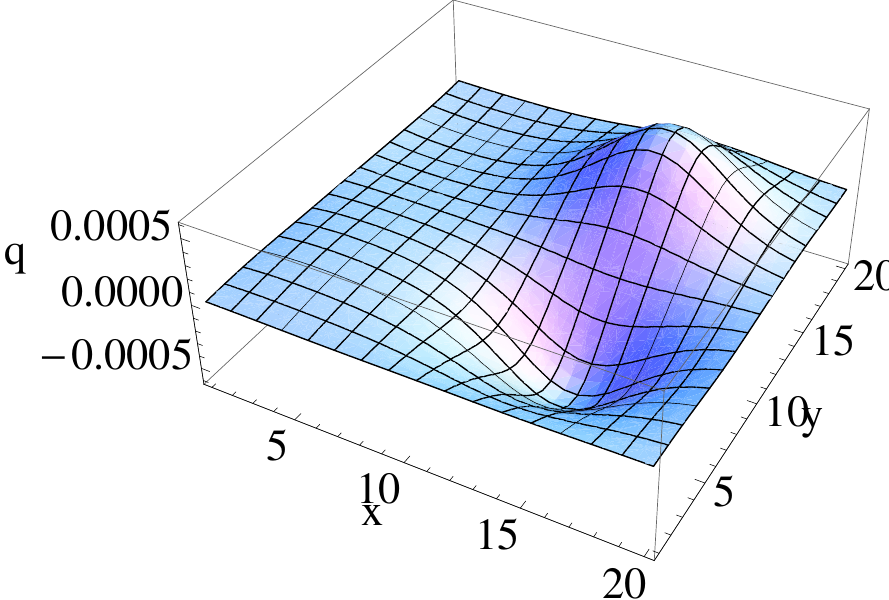}%
 c) \includegraphics[width=.31\textwidth]{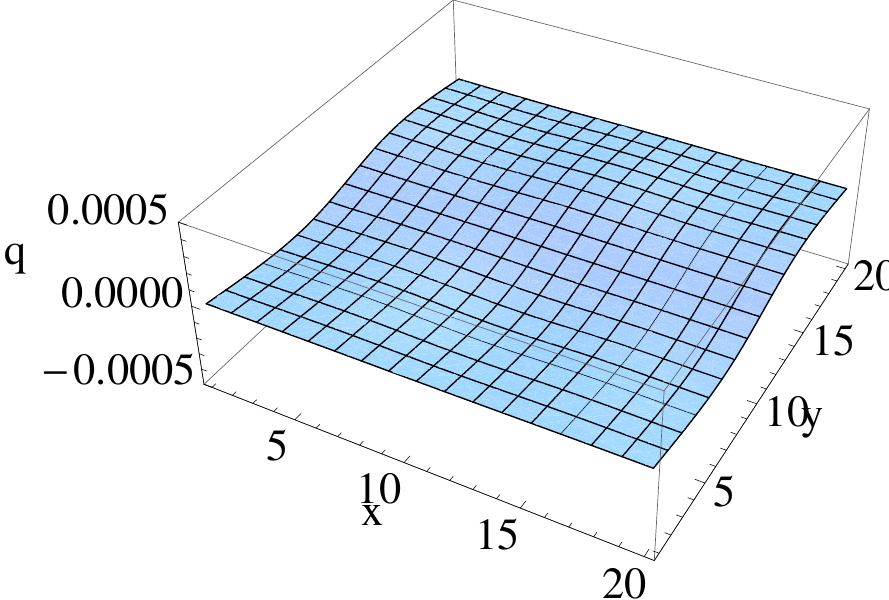}\\
\caption{For the light double-dyon-antidyon pair the reconstructed fermionic
topological charge densities are shown in a),b),c) corresponding to the
three boundary conditions mentioned in the text.}
\label{overdad}
\end{figure*}

Another example of an artificial dyon-antidyon system is a pair of two light
dyon-antidyon pairs formed out of the two light types (first and second) of the
same caloron solution as discussed above. For this aim we have placed
the constituents of the caloron at
\beqa
x_1=-1,  & ~~~ y_1=~~1,  & ~~~ z_1=0\nonumber \\
x_2=~1,  & ~~~ y_2=~~1,  & ~~~ z_2=0\\
x_3=~0,  & ~~~ y_3=-10,  & ~~~ z_3=0\nonumber
\eeqa
and those of the corresponding anticaloron at
\beqa
\bar x_1=-1,  & ~~~ \bar y_1=-1,  & ~~~ \bar z_1=0\nonumber \\
\bar x_2=~1,  & ~~~ \bar y_2=-1,  & ~~~ \bar z_2=0\\
\bar x_3=~0,  & ~~~ \bar y_3=10,  & ~~~ \bar z_3=0\nonumber,
\eeqa
respectively, and applied the same cut-and-paste procedure for the half
spaces $y>0$ and $y<0$ as before.

The pattern of near-zero eigenvalues of the overlap operator for the
extracted light double-dyon-antidyon pair is shown in \Fig{specdad}
for the three types of boundary conditions.
Now we observe a clear gap opening around zero for the third kind
of boundary conditions, while for the other ones near-zero eigenvalues occur.

The gluonic topological charge density,  as well as the set of monopole currents,
shows all dyons and antidyons, independent of their type, as can be seen in
\Fig{dad}a and \Fig{dad}b.
The Polyakov loop scatter plot for the (light) double-dyon-antidyon pair
(cf. Fig.~\ref{dad}c) is not a simple combination of Polyakov loop plots
for single dyon-antidyon pairs of a given type (compare with \Fig{tad}c),
but has a dispersed form due to the influence of dyons (antidyons)
of different type on each other.

For the same configuration the fermionic topological
charge densities are shown in \Fig{overdad}. As expected, only the
first and second type of fermionic boundary conditions visualize
the topological lumps of the respective light dyon-antidyon pairs.

\section{Results for the Yang-Mills ensemble}
\label{sec:ensemble_results}

In the following we will analyze the gluodynamics ensemble of 50 thermalized
configurations along the lines sketched above for the model dyon-antidyon
pairs. For identifying topological clusters of the lattice gauge fields with
the help of the low-lying spectrum of the overlap operator we used a fixed
number of 20 lowest modes always determined before any cooling or smearing 
was applied. In order to detect gluonic features of (anti)dyon excitations 
inside such clusters we employed four steps of over-improved
cooling~\cite{GarciaPerez:1993ki}. This amount of cooling changes (clarifies) 
the conformation of what we call the thermal monopole structure. The number 
of thermal monopoles was reduced by an approximate factor 2, and they became 
strictly static. Cooling beyond that stage kept the monopole number stable 
for a long period of cooling. Within four cooling steps, we did not 
completely match the topological profiles (gluonic and fermionic) as we did 
in our previous paper \cite{Ilgenfritz:2013oda} where we followed the concept
of an equivalent filtering as developed in
\cite{Bruckmann:2006wf,Ilgenfritz:2008ia,Bruckmann:2009vb}.

\begin{figure*}[!htb]
\centering
a)\hspace{0.3cm}\includegraphics[width=.37\textwidth]{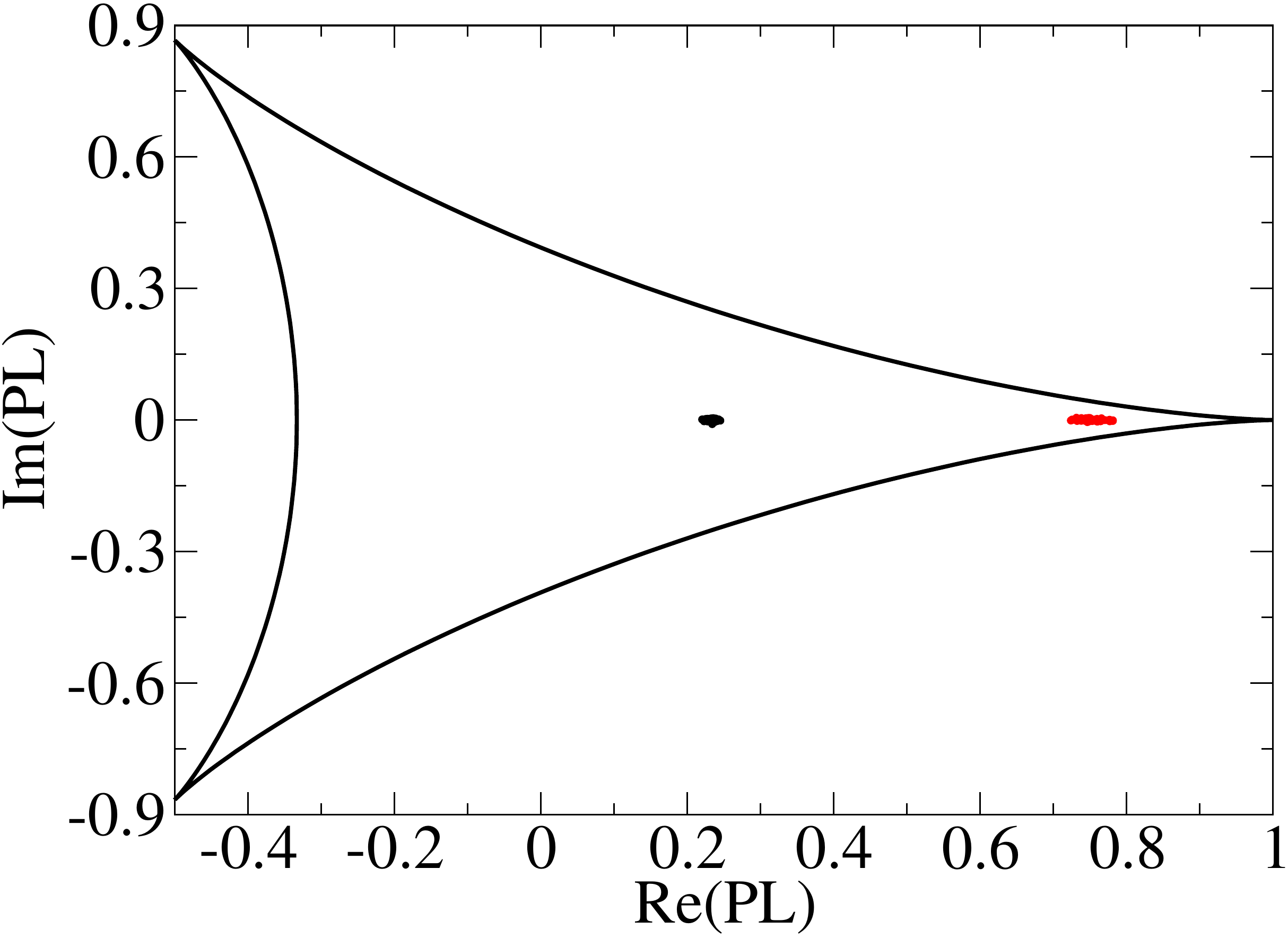}%
\hspace{0.3cm}
b)\hspace{0.3cm}\includegraphics[width=.37\textwidth]{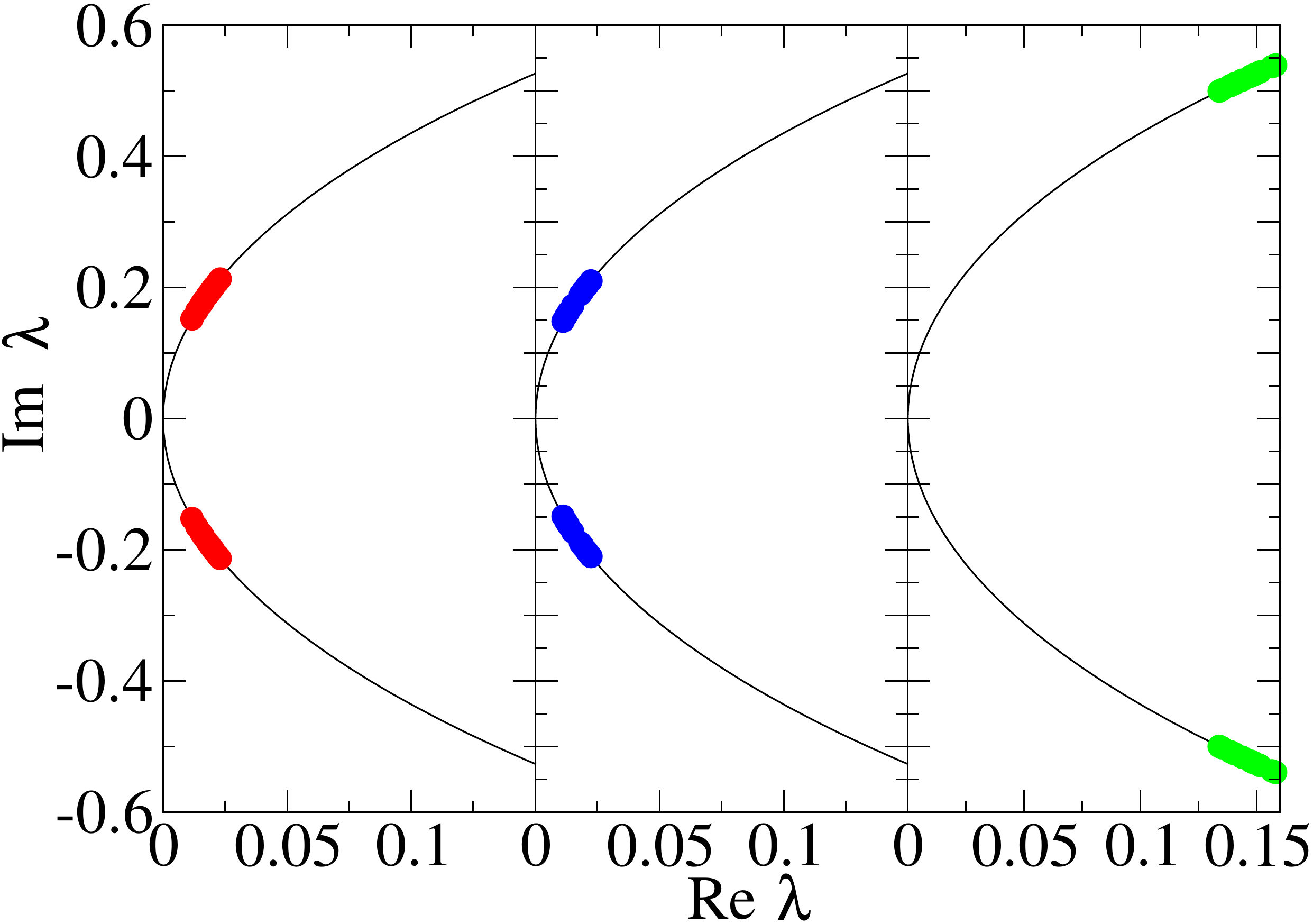}\\
\caption{
a) Scatter plots of the spatially averaged Polyakov loop $\overline L$
for 50 Monte Carlo generated configurations (shown in black symbols);
the right group (of red points) refers to the values
of $\overline L$ obtained after four steps of over-improved cooling,
b) overlap eigenvalues for one of these configurations under the three
boundary conditions.}
\label{pl204}
\end{figure*}

In \Fig{pl204}a we show a scatter plot of the spatially averaged Polyakov
loop $\overline L$ obtained from the ensemble of 50 generated configurations.
The (black) points concentrated around $ \overline L \simeq 0.24$ are belonging
to the Monte Carlo equilibrium configurations, while the shifted (red) points
(around $ \overline L \simeq 0.75$) correspond to the same configurations but
after the four steps of over-improved cooling.

It is clearly seen that the $Z(3)$-symmetry is spontaneously broken for the
equilibrium configurations as expected for temperatures above the critical one.
(Over-improved) cooling enhances this effect. Identifying the average
$\langle \overline L \rangle \simeq 0.24$ with the asymptotic holonomy
of an assumed dyon-antidyonic content of the gauge fields we conclude that
such a content would render dyons of first and second type lighter than
dyons of the third type. Therefore, we expect for equilibrium configurations
that dyons of the third (heavy) type will gain a smaller statistical weight.
(This differs from the situation of maximally nontrivial holonomy in the
confinement phase ($\langle \overline L \rangle \simeq 0$), where we expect
all the (anti)dyons occur with the same ``mass'' and statistical weight,
respectively.)

That such a splitting may happen is supported by the following observation.
The three eigenvalue spectra of the overlap operator obtained with the
three boundary conditions (for a typical configuration see \Fig{pl204}b)
look different. The third boundary condition,
which is the physical one in QCD with fermions, provides a much larger gap
than the others.  Thus, the spectra qualitatively resemble those observed
for the light double-dyon-antidyon pair (first- and second-type dyons) as
shown in \Fig{specdad}.
Therefore, we believe that light dyon-antidyon pairs (involving dyons of
the two light species) form the bulk of configurations in the deconfinement
phase and present further evidence below.
Concerning the topological charge density, we have applied the same cluster
analysis as in our previous paper \cite{Ilgenfritz:2013oda} with a variable
lower cutoff $q_{\rm cut}>0$ to analyse the density functions of
\Eq{eq:truncated_density} for thermal configurations describing the deconfined
phase. Let us repeat here the idea of the cluster algorithm.

In a first step - for each of the
three fermionic boundary conditions \Eq{eq:bc2} - the algorithm identifies the
points forming the interior of all clusters (the so-called ``topological
cluster matter'') defined by the condition $|q(x)| > q_{\rm cut}$.
The crucial second step is to enquire the connectedness between the lattice
points in order to form individual clusters out of this ``cluster matter''.
Neighbouring points with $|q(x)|$ above threshold and sharing the same sign of
the topological charge density are defined to belong to the same cluster.
The cutoff $q_{\rm cut}$ has been chosen such as to resolve the given
continuous distribution $q(x)$ into a maximal number of internally connected,
while mutually separated clusters. The cutoff value has been independently
adapted for each configuration.
As a result in the average the linear cluster size turns out approximately
$3.2 a \simeq 0.35$ fm.

We cannot exclude that this procedure might overestimate the number of
separately counted clusters by inclusion of too small objects with too
low density. But in any case, it allows to discover extended objects that
eventually can be qualified as (anti)dyons in the deconfined phase.
There are two conditions to make this interpretation in each case more 
likely: the local correlation with time-like Abelian monopoles in MAG 
and the occurence of nearly coinciding eigenvalues of the local
holonomy in the centers of all clusters. 

Thus, we have to inquire several criteria in order to enforce the evidence
for the dyonic nature of these clusters in the deconfined phase,
in the sense of being KvBLL caloron constituents.
In our previous work~\cite{Ilgenfritz:2013oda} we have concentrated on the
profile of the local Polyakov loop inside them,
which points towards the relative closeness of two (or three) eigenvalues
of the holonomy. Here additionally we use MAG monopoles as another feature
characterizing dyons. We have seen this in the artificial examples of
dyon-antidyon pairs considered in \Sec{sec:analytic_dyons}.

The removal of entropic monopole fluctuations
(as result of over-improved cooling as mentioned above)
renders all monopole loops static in temporal direction.
Moreover, it maximizes the number of time-like monopole currents
contained in topological clusters compared to the number of time-like
monopole currents present in the whole lattice.

This latter criterion has been decisive to determine the actual number of
sweeps of over-improved cooling (four). At this cooling stage the average
action for the given volume turned out equal to $S = 61.2(2) S_{\rm inst}$.
The (non-integer) gluonic topological charge $Q_{\rm glue}$ according
to \Eq{gluonic_Q} for each configuration was found to be equal to the
(integer) fermionic topological charge $Q_{\rm over}$ (given by the index of
the overlap operator \Eq{eq:OverlapDirac}) within 10\% accuracy.
In our ensemble of 50 configurations we found 43 configurations with
$Q_{\rm over}=0$ and 7 configurations with $~|{Q_{\rm over}}|=1$, which leads for
our temperature $~T = 1.5~T_d~$ and lattice volume to a rough estimate
of the (suppressed) topological susceptibility
$\chi_t = \langle Q_{\rm over}^2 \rangle / V \simeq (82~\mathrm{MeV})^4\,.$

\begin{figure}[!htb]
\centering
\hspace{0.3cm}
\includegraphics[width=.35\textwidth]{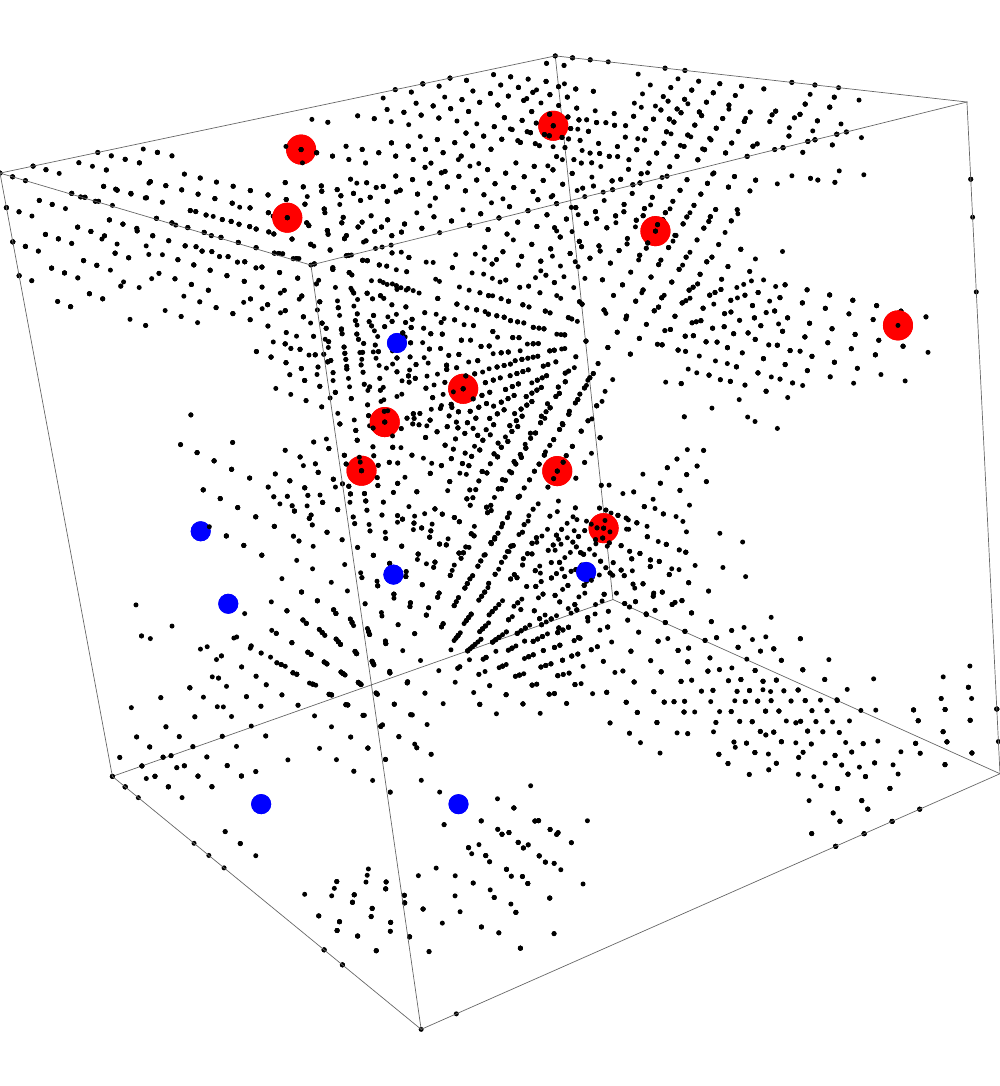}
\hspace{0.3cm}
\caption{The three-dimensional projection of points belonging to topological
clusters (black small points) and the location of static time-like monopole
loops after moderate over-improved cooling inside clusters (larger red spheres)
and outside clusters (small blue spheres) are shown for
one typical Monte Carlo generated gauge field configuration. }
\label{qmon}
\end{figure}

The three-dimensional projection of points belonging to topological clusters
and the location of the static monopole loops after four sweeps of
over-improved cooling steps are shown on Fig. \ref{qmon} for a typical
configuration.

All the data on the correlations between topological charge density and the
MAG monopoles is presented in Table I where the corresponding
data for cooled and original thermal configurations are shown for comparison.
Also the data obtained with only ten of the lowest overlap modes used for
determining the topological charge density are shown for comparison.
Let us note that the MAG Gribov copy effects measured by the difference between
results obtained with one and with 10 gauge copies amounts to
about 10\% for equilibrium configurations. For cooled configurations the results
do not differ within error bars.  In the following we will discuss results
obtained after cooling.

Our main results on the correlation of low-lying modes of the overlap Dirac
operator (as represented by the clusters of fermionic topological charge)
with the Abelian monopoles of MAG are as follows. Topological clusters occupy
about 16.8\% of the lattice volume, whereas topological clusters with static
MAG monopoles cover only 9.7\% of the lattice volume, but they contain about
35\% of MAG monopoles.
Inside topological clusters with MAG monopoles the latter are about 5 times
more dense than outside these clusters. These numbers become even more
pronounced if one counts not just the time-like monopole currents (dual links)
in topological clusters  but the numbers of thermal monopoles
piercing topological clusters. Then around 50\% of thermal monopoles
are piercing topological clusters.

We expect that the topological clusters detected with antiperiodic
boundary conditions (in our case with a real-valued average Polyakov loop)
can be viewed as related to heavy dyons which in the deconfinement
phase should become statistically suppressed because of their higher
action in comparison with the other constituents of a caloron at a holonomy
which is not maximally non-trivial.
We can estimate this suppression quantitatively
by measuring the abundance of MAG monopoles in topological clusters of
third type compared to those in topological clusters of first or second type.
We found after cooling and with twenty low-lying modes
the proportion $~~14 : 12 : 3.6~~$ (see the upper subtable).
Thus, the heavier caloron constituent clusters are really suppressed.

The following observations are also of interest. The average size of
clusters with magnetic monopoles is about four times larger than the
average size of clusters without magnetic monopoles,
while their number is approximately an order of magnitude smaller.
Clusters of third type (heavy dyons) are pierced just by one thermal
(static) monopole world line. In the average, only 2.5
time-like currents of monopole loops (out of 4 belonging to
a thermal monopole after cooling) are running inside these clusters.

In the case of topological clusters of first and second types
(light dyon candidates) in the average 3 time-like currents of monopole
loops (out of 4 belonging to a thermal monopole) are running inside
these clusters. Moreover, approximately 30\% of these clusters are
pierced even by two monopole loops. In order to understand this observation
one should take into account that clusters of the two light types occupy
a volume approximately twice as large as that of clusters of the third
type (identified as heavy dyons) and therefore, might overlap in space-time.
Unfortunately, to distinguish the monopoles (to make the intersections
one-to-one) is not a gauge-invariant concept.

\begin{table*}[ht]
\begin{center}
\vspace*{0.5cm}
Clusters obtained with 20 lowest overlap modes,  monopoles after cooling
\begin{tabular}{lllcccccc}
\hline
 Type of clusters & $V_{cl}$ & $V_{clmon}$ & $ N_{cl} $ & $N_{clmon}$ & $ N_{mon} $ & $N_{moncl}$
  & $ N_{loop} $ & $N_{loopcl}$ \\
\hline
3-d type (heavy) clusters  &$4.3(3)\%$ & $1.1(2)\%$ & $ 20(1) $ & $1.3(1)$ & $ - $ & $3.6(5)$& $ - $ & $1.4(2)$ \\
\hline
1-st type (light) clusters  &$8.5(6)\%$ & $5.4(6)\%$ & $ 25(1) $ & $3.7(2)$ & $ - $ & $14(1)$& $ - $ & $4.7(3)$ \\
\hline
2-nd type (light) clusters &$8.0(7)\%$ & $4.7(7)\%$ & $ 25(1) $ & $3.4(2)$ & $ - $ & $12(1)$& $ - $ & $4.3(3)$ \\
\hline
\hline
All clusters in total &$16.8(7)\%$ & $9.7(7)\%$ & $ 70(1) $ & $8.4(4)$ &
$ ~60(2)~/~64(2) $ & $21(1)$& $ ~15(1)~/~16(1) $ & $7.2(3)$ \\
\hline
\end{tabular}

Clusters obtained with 20 lowest overlap modes, monopoles before cooling\\
\begin{tabular}{lllcccccc}
\hline
 Type of clusters & $V_{cl}$ & $V_{clmon}$ & $ N_{cl} $ & $N_{clmon}$ & $ N_{mon} $ & $N_{moncl}$
  & $ N_{loop} $ & $N_{loopcl}$ \\
\hline
3-d type (heavy) clusters  &$4.3(3)\%$ & $2.2(2)\%$ & $ 20(1) $ & $4.3(3)$ & $ - $ & $12(1)$& $ - $ & $4.4(3)$ \\
\hline
1-st type (light) clusters  &$8.5(6)\%$ & $7.1(6)\%$ & $ 25(1) $ & $7.2(3)$ & $ - $ & $32(2)$& $ - $ & $11.8(7)$ \\
\hline
2-nd type (light) clusters &$8.0(7)\%$ & $6.5(6)\%$ & $ 25(1) $ & $7.0(3)$ & $ - $ & $29(2)$& $ - $ & $10.7(7)$ \\
\hline
\hline
All clusters in total &$16.8(7)\%$ & $13.3(7)\%$ & $ 70(1) $ & $18.5(6)$ &
$188(3)~/~210(4) $ & $55(2)$& $~32(1)~/~35(1) $ & $18(1)$ \\
\hline
\end{tabular}

Clusters obtained with only 10 lowest overlap modes,  monopoles after cooling\\
\begin{tabular}{lllcccccc}
\hline
 Type of clusters & $V_{cl}$ & $V_{clmon}$ & $ N_{cl} $ & $N_{clmon}$ & $ N_{mon} $ & $N_{moncl}$
  & $ N_{loop} $ & $N_{loopcl}$ \\
\hline
3-d type (heavy) clusters  &$4.4(4)\%$ & $1.3(2)\%$ & $ 13(1) $ & $1.3(1)$ & $ - $ & $3.6(4)$& $ - $ & $1.4(1)$ \\
\hline
1-st type (light) clusters  &$8(1)\%$ & $6(1)\%$ & $ 19(1) $ & $3.0(2)$ & $ - $ & $12(1)$& $ - $ & $4.2(3)$ \\
\hline
2-nd type (light) clusters &$10(1)\%$ & $8(1)\%$ & $ 18(1) $ & $2.7(2)$ & $ - $ & $12(1)$& $ - $ & $4.1(3)$ \\
\hline
\hline
All clusters in total &$18(1)\%$ & $13(1)\%$ & $ 50(1) $ & $7.0(3)$ &
                      $ ~~~~~~60(2)~~~~~~ $ & $21(1)$& $ ~~~~~~15(1)~~~~~~ $ & $6.9(3)$ \\
\hline
\end{tabular}
\label{tabdata}
\vspace*{0.5cm}
\caption{Results of the cluster analyse using low-lying overlap operator
modes with three kinds of boundary conditions, acc. to \Eq{eq:bc2}.
All numbers indicate averages per configuration. The pure statistical errors
are given in parentheses.
We denote with
$V_{\rm cl}$     - the volume fraction occupied by all topological clusters,
$V_{\rm cl~mon}$ - the volume fraction occupied by clusters containing
                   time-like magnetic monopoles,
$N_{\rm cl}$     - the number of all clusters per configuration,
$N_{\rm cl~mon}$ - the number of clusters containing time-like
                   magnetic monopoles,
$N_{\rm mon}$    - the overall number of dual timelike links
                   carrying monopole currents,
$N_{\rm mon~cl}$ - the number of dual timelike links with monopole currents
                   found inside topological clusters,
$N_{\rm loop}$   - the overall number of  thermal monopoles,
$N_{\rm loop~cl}$ - the number of timelike magnetic current loops piercing
                   topological clusters.
The effect of Gribov copies (see the text) on $N_{\rm mon}$
and $N_{\rm loop}$ for cooled and original configurations is indicated
in the last lines of the upper two subtables by '.. / ..'.
}
\end{center}
\end{table*}

Although the pattern of Polyakov loops becomes highly modified by cooling
in the deconfined phase, it is possible to point out a correlation between
the Polyakov loop on one side and monopoles, respective clusters of
topological charge on the other.

First, let us compare the distributions of the minimal distance between
eigenvalues of the local holonomies for all lattice sites and for sites
carrying thermal monopoles. From analytical caloron solutions
and from our artificial semianalytic configurations
(see Figs.  \ref{tad}c and \ref{dad}c) we know that in the center of
a topological dyon cluster with a magnetic monopole the local holonomy
has at least two identical eigenvalues. This means that the local Polyakov
loop takes a value on one of the three sides of the Polyakov triangle
(see the Appendix in Ref.~\cite{Ilgenfritz:2013oda}).

\begin{figure}[!htb]
\centering
\vspace{1cm}
\includegraphics[width=.37\textwidth]{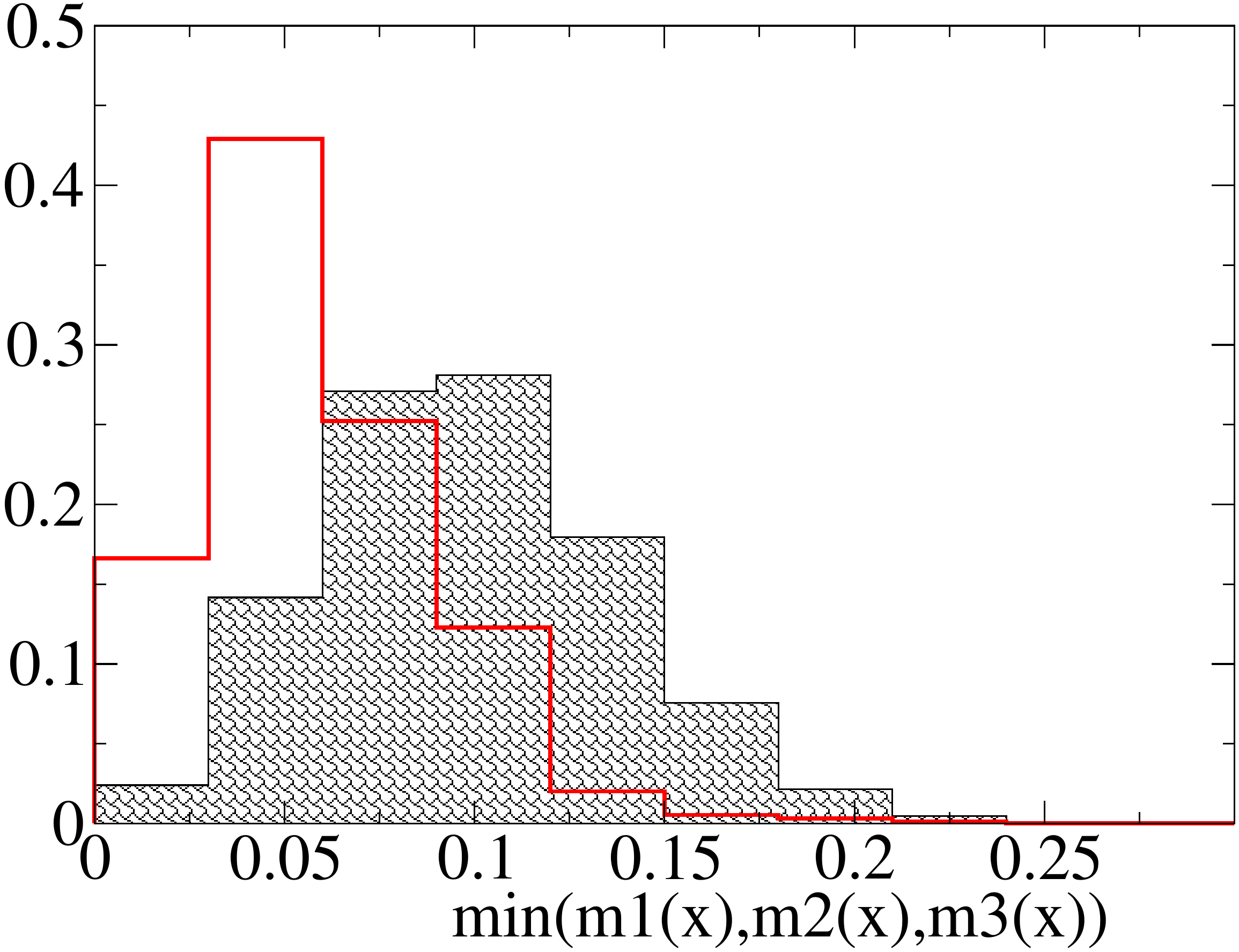}%
\caption{
For all lattice sites (shaded histogram) and for all cubes where thermal
monopoles are located (open red histogram) the
distributions with respect to the minimal distance
$\min(m_1(\vec{x}),m_2(\vec{x}),m_3(\vec{x}))$ between the Polyakov loop
and one of the boundaries of the Polyakov triangle are shown. In the case of
a monopole the minimum is taken also among the 8 corners of the
three-dimensional cube containing that monopole.
}
\label{mmindistr}
\vspace{0.2cm}
\end{figure}

We quantify the closeness of a Polyakov loop value to the boundary
of this triangle by the minimal
distance $\mbox{min}\{ m_1(\vec{x}), m_2(\vec{x}), m_3(\vec{x})\}$,
where the $m_i(\vec{x})$ are defined as the differences
between the three eigenvalues $\mu_i(\vec{x})$ of the local
holonomy according to Eqs. (\ref{eq:localholonomy}) and
(\ref{eq:localholonomy_evs}))
$$
m_i(\vec{x})=|\mu_{i+1}(\vec{x})-\mu_{i}(\vec{x})|,~i=1,2,3,~
\mu_4(\vec{x}) \equiv \mu_1(\vec{x})\,.
$$
The two distributions with respect to the minimal distance are shown in
\Fig{mmindistr} and tell that the local Polyakov loop at sites with thermal
monopoles tend to be located closer to the boundary of the Polyakov triangle
than for all lattice sites.

\begin{figure*}[!htb]
\centering
a)\hspace{0.3cm}\includegraphics[width=.37\textwidth]{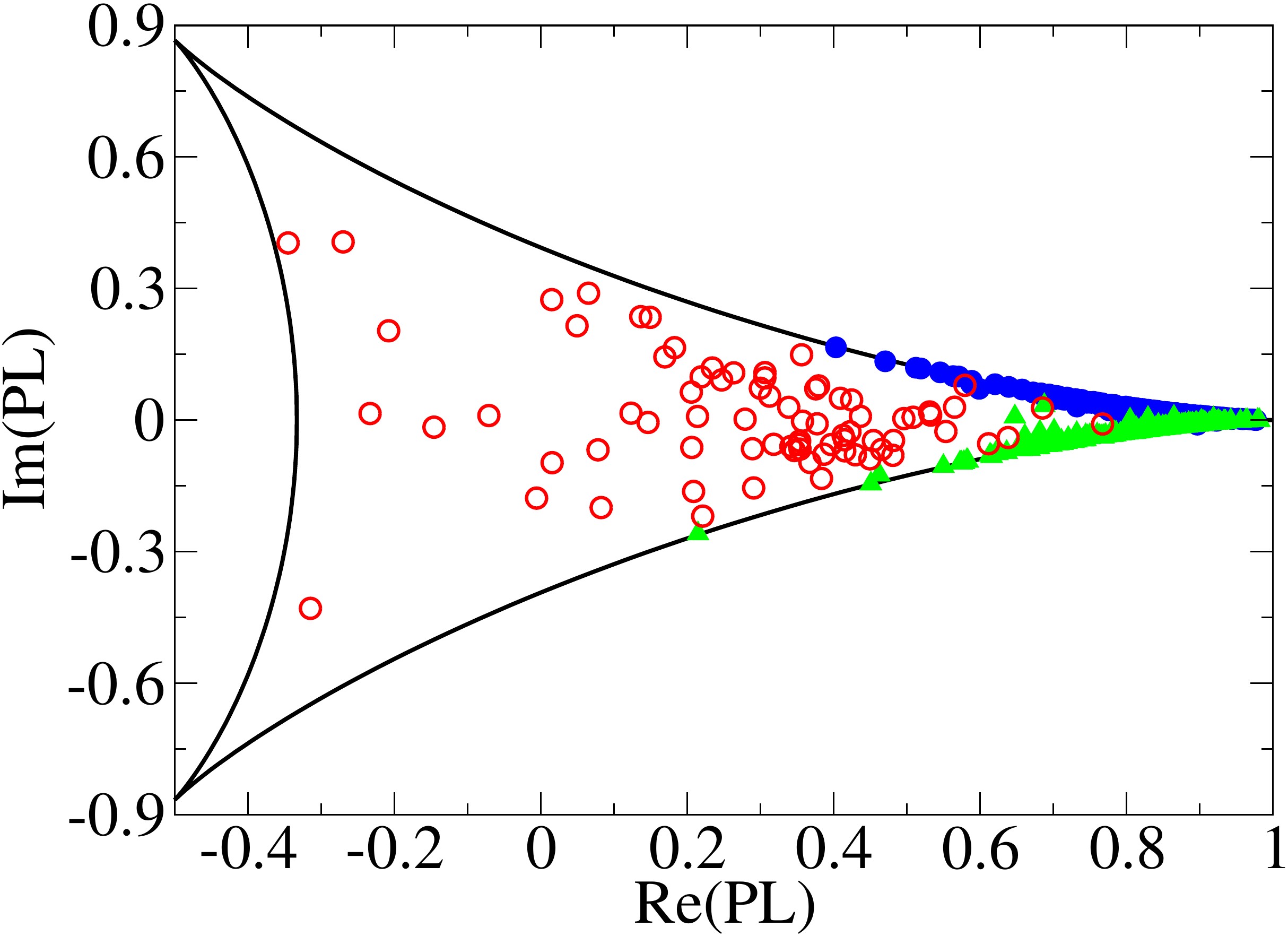}%
\hspace{0.3cm}
b)\hspace{0.3cm}\includegraphics[width=.37\textwidth]{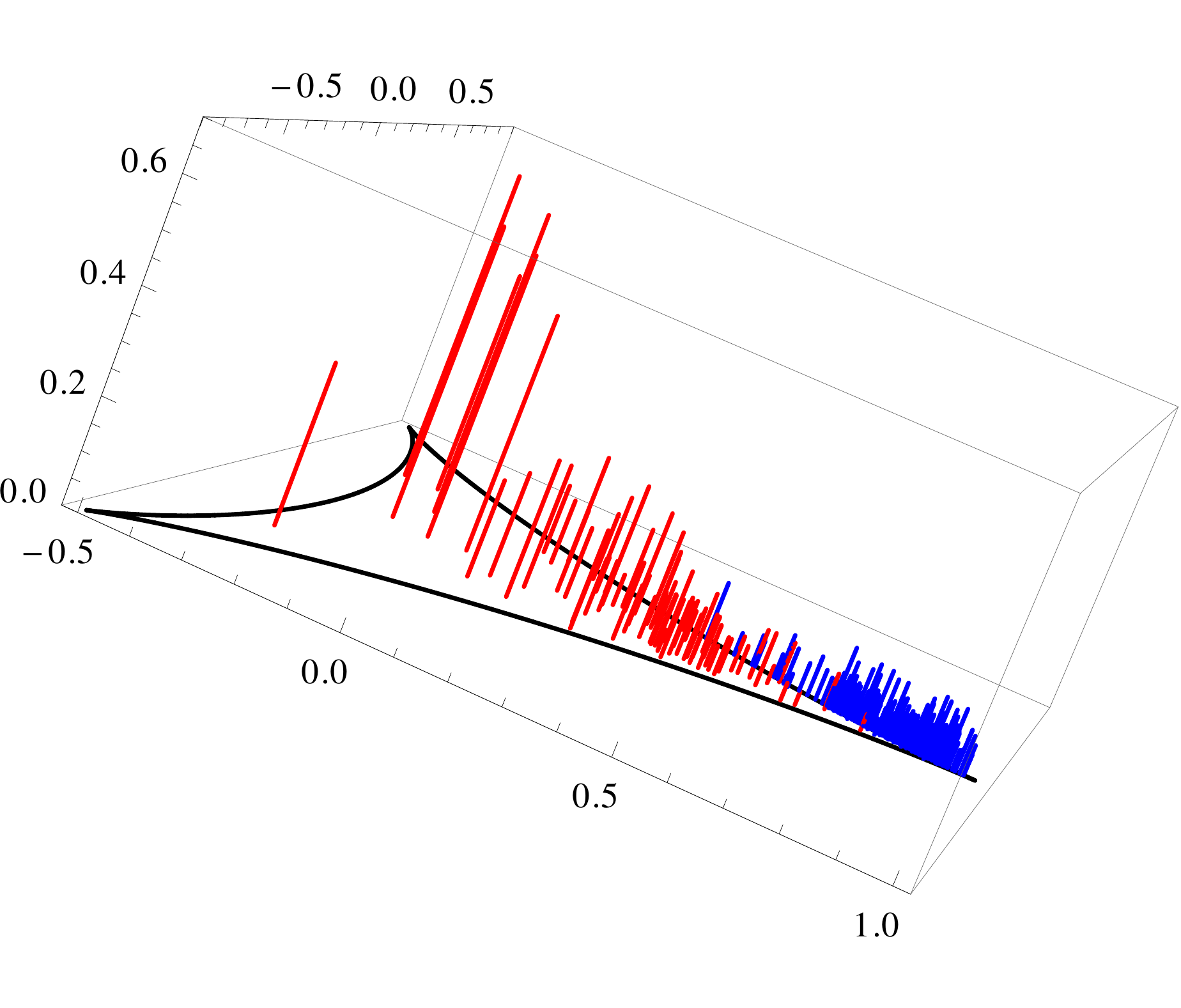}
\vspace*{0.7cm}
\caption{
a) Scatter plots of Polyakov loop $PL$ (after 4 steps of cooling)
in clusters selected to contain monopoles. The clusters are separated
according to the type of boundary condition for the overlap near-zero modes.
For clusters of first type the Polyakov loop is shown by green triangles,
for clusters of second type - by blue filled circles,
for clusters of third type - by red open circles,
b) The maximum of the topological charge density  inside the respective cluster
is additionally shown in respective color (for second and third type clusters
only).}
\label{plq204}
\vspace{1.5cm}
\end{figure*}

Second, we show the scatter plot of Polyakov loops measured (after cooling)
in the centers of those clusters which are associated with magnetic monopoles.
Since the clusters are labelled by one of the three boundary conditions for
the fermionic modes (used to define the fermionic topological charge density),
the scatter plot over the Polyakov triangle \Fig{plq204}a~ shows the
different regions of population. 
There is a tendency of the Polyakov loop in the centers of topological clusters
of the two light kinds to populate two sides of the Polyakov triangle beginning
from the trivial Polyakov loop $L \approx (1.0,0.0)$. 
Compared with the results before cooling, the population has moved 
closer towards the trivial Polyakov loop and towards the periphery, thereby 
improving the (approximate) degeneracy of two eigenvalues of the local holonomy.
From \Fig{plq204}a~ we see also that clusters of the third type (which are heavy) 
are less abundant and distributed over most of the Polyakov triangle.
Cooling has moved part of them towards the trivial Polyakov loop, too,
but others are still differing strongly from trivial holonomy.

Finally, if one extends the scatter plots by a third dimension representing
the maximal absolute value of topological charge density of the corresponding
clusters by spikes (see \Fig{plq204}b ) one observes the clusters of first and
second type to have negligible topological charge, while the clusters of third
type may carry noticeable topological charge (deserving the name heavy clusters).
There is a tendency of heavy clusters to have a Polyakov loop opposite to
the trivial one, $L \approx (1.0,0.0)$.

In conclusion, the Polyakov loop characteristics of the ``light plus heavy
dyonic picture'' for the clusters of topological charge in the deconfined
phase is clearly visible after a slight cooling of the configurations.

\section{Conclusions}
\label{sec:conclusions}

For $SU(3)$ gluodynamics we have discussed the signatures of dyonic
topological excitations of thermal lattice gauge fields generated in the
deconfinement phase. We have chosen a temperature value $T=1.5~T_c$, i.e.
well above the critical one. Under the assumption that (anti)dyon excitations
become really relevant in the sense proposed by Diakonov and Petrov 
\cite{Diakonov:2007nv} we suppose them to be related to the constituents
of KvBLL calorons \cite{Kraan:1998pm,Kraan:1998sn,Lee:1998bb} with an 
asymptotic holonomy determined by the average Polyakov loop 
(always taken in the real sector of the Polyakov triangle)
which is then clearly different from zero (i.e. different from maximally
non-trivial holonomy in the confinement phase). In this case the three
monopole (dyon) constituents of KvBLL calorons are known to differ
with respect to their masses or summed topological charges, the latter
being directly related to the eigenvalues of the asymptotic holonomy.
Then it is natural to conjecture that the heavy kind of dyons will be
statistically suppressed compared with their light type. This means also
that full KvBLL calorons should become rare excitations, too.
It was our task to provide a numerical evidence for this semiclassical-like 
dyon picture.

In order to find signatures of distinct light and heavy (anti)dyon pairs
we have first constructed classical model configurations
from KvBLL (anti)caloron solutions with the help of an appropriate
cut-and-paste procedure. For these configurations we checked the fermion‎ic
overlap eigenvalue spectrum and visualized them with several local observables:
\begin{itemize}
\item the gluonic topological density,
\item the fermionic topological density filtered with the low-lying modes
      of the overlap operator and determined with a set of three different
      time-like boundary conditions, such that each boundary condition
      extracts just one dyon-type,
\item local values of the Polyakov loop as corresponding to the local
      holonomies for which the degeneracy of eigenvalues are pointing
      to the positions of the dyon constituents,
\item the Abelian monopole currents in the maximally Abelian gauge.
\end{itemize}
For the examples of a heavy dyon-antidyon pair and for a light
double-dyon-antidyon pair we produced a very clear pattern to be
qualitatively compared with that of topological clusters of Monte Carlo
generated quantum gauge fields.

Such topological clusters were then established by filtering with
20 low-lying modes of the overlap Dirac operator by employing the same
three boundary conditions. Additionally we subjected the lattice fields
to a few (overimproved) cooling steps after which a similar pattern of
clusters occurs with the gluonic topological charge distribution.
With and without cooling we looked for the behavior of the spatially
averaged as well as the local distributions of the Polyakov loop
(as well as its local holonomies) and searched for MAG monopole
currents.

First of all - depending on the boundary conditions - we mostly found
eigenvalue spectra similar to those produced by light dyon-antidyon pairs
and rare cases telling about heavy dyon-antidyon pairs.
Moreover, we found clear correlations of the topological clusters
with thermal monopoles as well as with lattice sites, where the
local holonomy has close-to-degenerate eigenvalues.

All this points to an interpretation in terms of mostly light -
with only a dilute admixture of heavy - (anti)dyon excitations of the
KvBLL type.

Moreover, our findings resemble very much to what we found earlier
in the $SU(2)$ case~\cite{Ilgenfritz:2006ju,Bornyakov:2007fm,Bornyakov:2008im}
where in the deconfinement phase the dominance of light dyon
constituents was seen, too.

\noi
{\bf Acknowledgments} \\
B.V.M. appreciates the support of Humboldt-University
Berlin where the main part of the work was finalized.
E.-M.I. and M.M.-P. acknowledge financial support by the
Heisenberg-Landau Program of BLTP at JINR Dubna.
V.G.B. has been supported by the grant RFBR 13-02-01387a.

\bibliographystyle{apsrev}

\end{document}